\documentclass[12pt]{iopart}
\usepackage{iopams}
\usepackage{slashed}
\usepackage{epsfig}

\def\np#1#2#3{Nucl.~Phys.~{\bf B {#1}} ({#2}) #3}

\def\pr#1#2#3{Phys.~Rev.~{\bf D {#1}} ({#2}) #3}

\begin{document}

\setcounter{footnote}{0}
\def\a{\alpha}
\def\b{\beta}
\def\c{\chi}
\def\d{\delta}
\def\e{\epsilon}
\def\f{\phi}
\def\g{\gamma}
\def\h{\eta}
\def\i{\iota}
\def\j{\psi}
\def\k{\kappa}
\def\l{\lambda}
\def\m{\mu}
\def\n{\nu}
\def\o{\omega}
\def\p{\pi}
\def\q{\theta}
\def\r{\rho}
\def\s{\sigma}
\def\t{\tau}
\def\u{\upsilon}
\def\x{\xi}
\def\z{\zeta}
\def\D{\Delta}
\def\F{\Phi}
\def\G{\Gamma}
\def\J{\Psi}
\def\L{\Lambda}
\def\O{\Omega}
\def\P{\Pi}
\def\Q{\Theta}
\def\S{\Sigma}
\def\U{\Upsilon}
\def\X{\Xi}

\def\ve{\varepsilon}
\def\vf{\varphi}
\def\vr{\varrho}
\def\vs{\varsigma}
\def\vq{\vartheta}
\def\dg{\dagger}                                     
\def\ddg{\ddagger}                                   
\def\wt#1{\widetilde{#1}}                    
\def\mt{\widetilde{m}_1}
\def\mti{\widetilde{m}_i}
\def\rt{\widetilde{r}_1}
\def\mtt{\widetilde{m}_2}
\def\mttt{\widetilde{m}_3}
\def\rtt{\widetilde{r}_2}
\def\mb{\overline{m}}
\def\VEV#1{\left\langle #1\right\rangle}        
\def\be{\begin{equation}}
\def\ee{\end{equation}}
\def\ds{\displaystyle}
\def\ra{\rightarrow}

\def\bea{\begin{eqnarray}}
\def\eea{\end{eqnarray}}
\def\NO{\nonumber}
\def\Bar#1{\overline{#1}}

\title{The minimal scenario of leptogenesis}

\author{Steve Blanchet}
\address{Institut de Th\'eorie des Ph\'enom\`enes Physiques,
\'Ecole Polytechnique F\'ed\'erale de Lausanne, CH-1015 Lausanne, Switzerland}

\author{Pasquale Di Bari}
\address{School of Physics and Astronomy, University of Southampton, Southampton SO17 1BJ, UK}

\begin{abstract}
We review the main features and results of thermal leptogenesis
within the type I seesaw mechanism,  the minimal extension of
the Standard  Model explaining  neutrino masses and mixing.
After  presenting the simplest approach,  the vanilla scenario, we discuss
various important  developments in recent years, such as the inclusion of lepton and heavy
neutrino  flavour effects,  a description beyond a hierarchical heavy neutrino mass spectrum
and an improved kinetic description within the density matrix and the closed-time-path formalisms.
We also discuss how
leptogenesis can ultimately represent an important phenomenological tool
to test the seesaw mechanism and the underlying model of new physics.
\end{abstract}

\maketitle

\section{Introduction: the double side of leptogenesis}

A successful model of baryogenesis cannot be
realised within the Standard Model (SM) and, therefore,
the observed matter-antimatter asymmetry of the Universe
can be regarded as an evidence for new physics beyond the SM.

The discovery of neutrino masses and mixing in neutrino oscillation
experiments in 1998 \cite{superk} has for the first time shown directly,
in particle physics experiments, that the SM is indeed incomplete, since it 
strictly predicts, in the limit of infinite cutoff, 
that neutrinos are massless and, therefore, cannot oscillate.

This discovery has  greatly raised the interest for
leptogenesis \cite{Fukugita:1986hr,reviews},  a model of  baryogenesis that is a cosmological
consequence of the most  popular way to extend the SM in order to
explain why neutrinos are massive but at the same time much lighter than all the
other fermions: the seesaw mechanism~\cite{seesaw}.
As a matter of fact, leptogenesis  realises a highly non-trivial link
between two completely independent experimental observations:
the absence of primordial antimatter in the observable Universe and the
observation that neutrinos mix and (therefore)  have masses.
In fact, leptogenesis has a naturally built-in double-sided nature.
On one side, it describes a very early stage in the history of the Universe
characterised by temperatures $T_{\rm lep} \gtrsim 100\,{\rm GeV}$,
much higher than those probed by Big Bang Nucleosynthesis, $T_{\rm BBN} \sim (0.1$--$1)\,{\rm MeV}$;
on the other side, it complements low-energy neutrino experiments providing
a completely independent phenomenological tool
to test models of new physics embedding the seesaw mechanism.

In this article we review the main features and results of leptogenesis.
Let us give a brief outline.
In Section 2 we present the status of low-energy neutrino experiments measuring
neutrino masses and mixing parameters, and we introduce the seesaw mechanism, which provides an elegant framework to explain them.
In Section 3 we discuss the vanilla leptogenesis scenario.
In Section 4 we show the importance of accounting for flavour effects for a correct calculation of the final asymmetry. In Section 5 we discuss the density matrix formalism which properly takes into account decoherence effects, which are
crucial to describe the transition from a one-flavoured regime to a fully flavoured regime.
In Section 6 we relax the assumption of hierarchical right-handed neutrino mass spectrum, and discuss how the asymmetry can be calculated in the degenerate limit.
In Section 7 we discuss different ways to improve the kinetic description
beyond the density matrix formalism.
In Section 8 we discuss other effects (thermal corrections, spectator processes, scatterings)
that have been considered and that can give in some cases important corrections.
In Section 9 we show how leptogenesis represents an important guidance to test models of new physics.
Finally, in Section 10 we conclude outlining the prospects to test leptogenesis in future years.

\section{Neutrino masses and mixing}

Neutrino oscillation experiments have established two fundamental
properties of neutrinos. The first one is that the neutrinos mix.
This means that the neutrino weak eigenfields
$\nu_{\alpha}$ ($\alpha=e,\mu,\tau$)
do not coincide with the neutrino mass eigenfields $\nu_i$ ($i=1,2,3$) but
are obtained applying to them a unitary transformation described
by the $(3\times 3)$  leptonic mixing matrix $U$,
\be
\nu_{\alpha} = \sum_{i} \, U_{\alpha i}\,\nu_i \, .
\ee
The leptonic mixing matrix is usually parameterised in terms of  6 physical parameters,
three mixing angles, $\theta_{12}$,  $\theta_{13}$ and $\theta_{23}$, and three phases,
two Majorana phases, $\rho$ and $\sigma$ and one Dirac phase $\delta$,
\be\label{Umatrix} \fl
U=\left( \begin{array}{ccc}
c_{12}\,c_{13} & s_{12}\,c_{13} & s_{13}\,e^{-{\rm i}\,\d} \\
-s_{12}\,c_{23}-c_{12}\,s_{23}\,s_{13}\,e^{{\rm i}\,\d} &
c_{12}\,c_{23}-s_{12}\,s_{23}\,s_{13}\,e^{{\rm i}\,\d} & s_{23}\,c_{13} \\
s_{12}\,s_{23}-c_{12}\,c_{23}\,s_{13}\,e^{{\rm i}\,\d}
& -c_{12}\,s_{23}-s_{12}\,c_{23}\,s_{13}\,e^{{\rm i}\,\d}  &
c_{23}\,c_{13}
\end{array}\right)
\, {\rm diag}\left(e^{i\,\rho}, 1, e^{i\,\sigma}
\right)\, ,
\ee
where $s_{ij} \equiv \sin\theta_{ij}$ and $c_{ij}\equiv\cos\theta_{ij}$.
A global analysis \cite{Fogli:2011qn} of all existing neutrino data, prior to the results
of a non-vanishing $\theta_{13}$ from short baseline reactors,  gives
$\theta_{12}=34^{\circ}\pm 1^{\circ}$ for the solar mixing angle,
$\theta_{23}= 40.4{^{\circ}}^{+4.6^{\circ}}_{-1.8^{\circ}}$
for the atmospheric mixing angle, and $\theta_{13}=9.0^{\circ}\pm 1.3^{\circ}$
for the reactor mixing angle, where the latter is mainly dominated by
the evidence for a non-vanishing $\theta_{13}$ found by the T2K experiment \cite{T2K}
that confirmed previous hints \cite{hints}.
Recently, the Daya Bay and Reno short-baseline reactor neutrino experiments
respectively found $\theta_{13}= 8.8^{\circ}\pm 0.8^{\circ}\pm 0.3^{\circ}$ \cite{dayabay},
and $\theta_{13}=9.8^{\circ}\pm 0.6{^{\circ}}\pm 0.85^{\circ}$ \cite{reno}, confirming, at more than $5\s$,
the previous results.

The second important property established by neutrino oscillation experiments
is that neutrinos are massive. More specifically, defining the three neutrino masses
in a way that $m_1 \leq m_2 \leq m_3$, neutrino oscillation experiments
measure two mass-squared differences that we can indicate with
$\Delta m^2_{\rm atm}$ and $\Delta m^2_{\rm sol}$ since historically the first
one has been first measured in atmospheric neutrino experiments and the second one in
solar neutrino experiments.
 Two options are currently allowed by previous
experiments. A first option is  `normal ordering' (NO) and in this case
\be
m_3^2 - m^2_2 = \Delta m^2_{\rm atm}  \hspace{5mm} \mbox{\rm and}
\hspace{5mm}
m_2^2 - m^2_1 = \Delta m^2_{\rm  sol} \, ,
\ee
while a second option is represented by `inverted ordering' (IO) and in this case
\be
m_3^2 - m^2_2 = \Delta m^2_{\rm sol}  \hspace{5mm} \mbox{\rm and} \hspace{5mm}
m_2^2 - m^2_1 = \Delta m^2_{\rm  atm} \,  .
\ee
It is convenient to introduce the atmospheric neutrino mass scale
$m_{\rm atm} \equiv \sqrt{\Delta m^2_{\rm atm}+\Delta m^2_{\rm  sol}}
= (0.049\pm 0.001)\,{\rm eV}$ and the solar neutrino mass scale
$m_{\rm sol} \equiv \sqrt{\Delta m^2_{\rm  sol}}=(0.0087 \pm 0.0001)\,{\rm eV}$
\cite{Fogli:2011qn}.

The measurements of $m_{\rm atm}$ and $m_{\rm sol}$
are not sufficient to fix all three neutrino masses.
If we express them in terms of the lightest neutrino
mass $m_1$ we can see from Fig.~1 that while
 $m_2 \geq m_{\rm sol}$ and $m_3 \geq m_{\rm atm}$, the lightest
neutrino mass can be arbitrarily small implying that the lightest neutrino
could  be even massless.

The lower limits for $m_2$ and $m_3$
are saturated when $m_1 \ll m_{\rm sol}$. In this case one has hierarchical
neutrino models, either normal and in this case $m_2 \simeq m_{\rm sol}$
and $m_3\simeq m_{\rm atm}$, or inverted, and in this case
$m_2\simeq \sqrt {m^2_{\rm atm}-m^2_{\rm sol}} \simeq m_3\simeq m_{\rm atm}$.
On the other hand, for $m_1\gg m_{\rm atm}$ one obtains the limit of quasi-degenerate neutrinos
when all three masses can be arbitrarily close to each other.

However, the lightest neutrino mass is upper bounded by absolute neutrino mass scale
experiments. Tritium beta decay experiments \cite{Kraus:2004zw}
place an upper bound on the effective electron  neutrino mass
$m_{\nu_e} \lesssim 2\,{\rm eV}$ ($95 \%$ C.L.) that translates
into the same upper bound on $m_1$.  This is derived from model-independent
kinematic considerations that apply independently of whether neutrinos have a Dirac or
Majorana nature.

Neutrinoless double beta decay ($0\nu\b\b$) experiments
place a more stringent upper bound on the effective $0\nu\b\b$ Majorana neutrino mass,
$m_{ee}\lesssim (0.34$--$0.78)\,{\rm eV}$\,($95\%$ C.L.) as obtained by the CUORICINO
experiment~\cite{bb0n}
\footnote[1]{At $90\%$~C.L. the CUORICINO result is $m_{ee}\lesssim (0.27$--$0.57)\,{\rm eV}$. For comparison,
the bound from the Heidelberg-Moscow experiment  is  $m_{ee}\lesssim (0.21$--$0.53)\,{\rm eV}$ ($90\%$~C.L.) 
\cite{HDM} while the EXO-200 experiment has recently found the upper 
bound $m_{ee}\lesssim (0.14$--$0.38)\,{\rm eV}$ ($90\%$~C.L.) \cite{EXO}.}.
It translates into the following upper bound on $m_1$~\cite{rodejohann}:
\be \fl
m_1 \leq m_{ee}/(\cos 2\theta_{12}\,\cos^2 \theta_{13}-\sin^2\theta_{13}) 
\lesssim 3.45 \, m_{ee} \lesssim (1.2 \textrm{--}2.7)\,{\rm eV}  
\hspace{3mm} (95\%\, {\rm C.L.})   \,  .
\ee
Here, the wide range is due to
theoretical uncertainties in the calculation
of the involved nuclear matrix elements.  However, this upper bound applies
only if neutrinos are of Majorana nature, which is the relevant case for us, since
the seesaw mechanism predicts Majorana neutrinos.
\begin{figure}
\begin{center}
\begin{minipage}{150mm}
\begin{center}
\centerline{\psfig{file=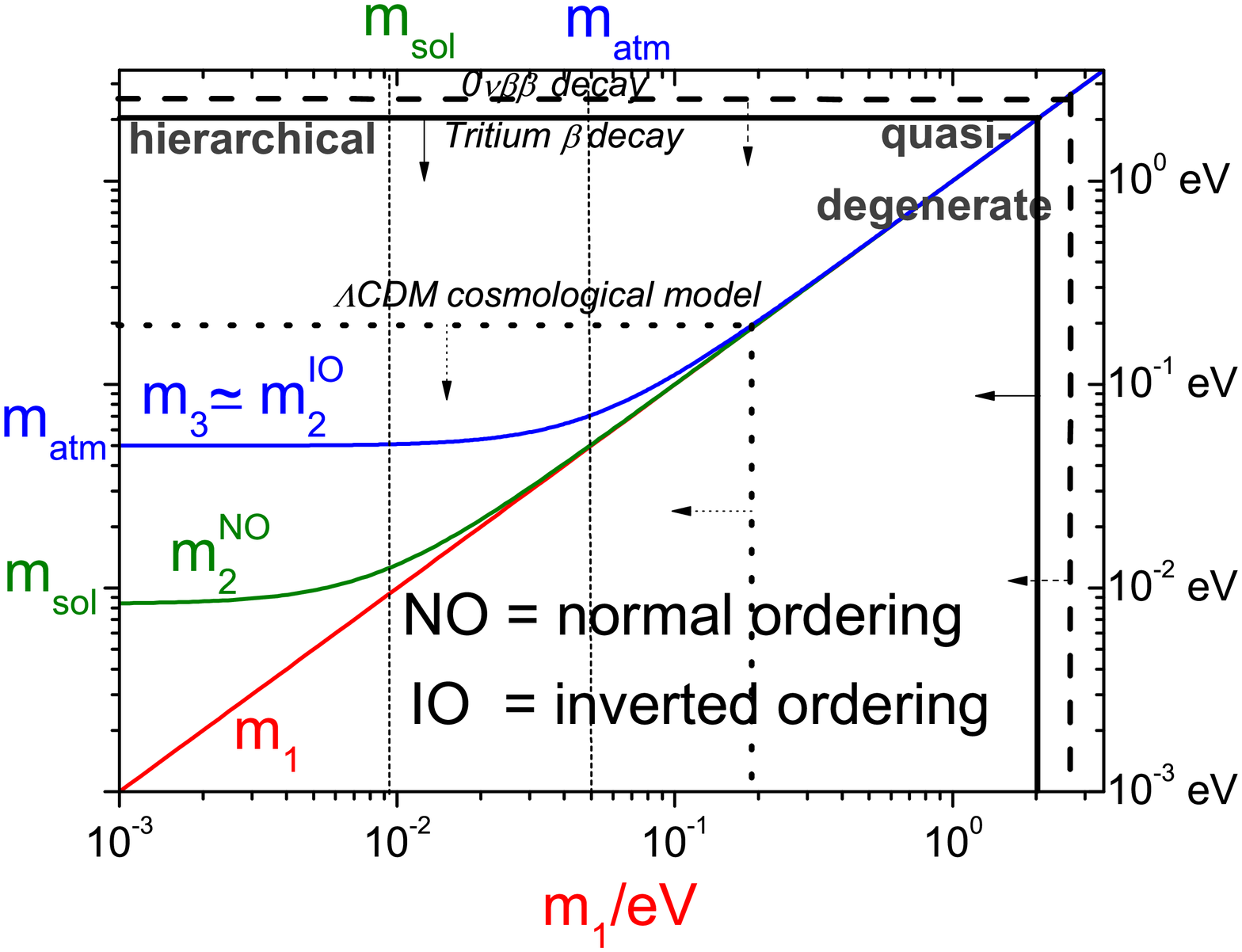,height=8cm,width=11cm}}
\caption{Neutrino masses $m_i$ versus the lightest neutrino mass $m_1$.
The three upper bounds (at $95\%$ C.L.)  discussed in the body text
from absolute neutrino mass scales phenomenologies are also indicated.}
\end{center}
\end{minipage}
\end{center}
\end{figure}

From cosmological observations,
within the $\L$CDM model, one obtains a very stringent upper bound on the
sum of the neutrino masses \cite{WMAP7}
\be\label{WMAP7}
\sum_i\, m_i \lesssim 0.58\,{\rm eV} \hspace{10mm} (95\% \, {\rm C.L.})\, .
\ee
This translates into the
most stringent upper bound that we currently have on the
lightest neutrino mass $m_1 \lesssim 0.19\,{\rm eV}$ ($95\%$ C.L.), an upper bound that
almost excludes quasi-degenerate neutrino models
\footnote[2]{For a more general discussion of neutrino mass bounds, 
in particular on theoretical assumptions and uncertainties, see for example 
\cite{rodejohann,giunti,pastor}.}.

A minimal extension of the SM, able to explain not only why neutrinos
are massive but also why they are much lighter than all the other massive fermions,
is represented by the seesaw mechanism \cite{seesaw}.
In the minimal type I version, one adds
right-handed neutrinos $N_{Ri}$ to the SM lagrangian with Yukawa couplings $h$
and a Majorana mass term that violates lepton number
\be\label{lagrangian}
\mathcal{L}  =  \mathcal{L}_{\rm SM}
+{\rm i} \, \overline{N_{iR}}\g_{\m}\partial^{\m} N_{iR} -
\overline{\ell_{\a L}}\,h_{\a i}\, N_{iR}\, \tilde{\F} -
 {1\over 2}\,M_i \, \overline{N_{iR}^c} \, N_{iR} +h.c. \nonumber
\ee
For definiteness we will consider the case of three RH neutrinos ($i=1,2,3$).
This is the most attractive case corresponding to have one RH neutrino for each generation,
as for example nicely predicted by $SO(10)$ grand-unified  models.
Notice, however, that all current
data from low-energy neutrino experiments are also consistent
with a more minimal model with only two RH neutrinos that will be discussed in Section 9.

After spontaneous symmetry breaking, a Dirac mass term $m_D=v\,h$
is generated by the  Higgs vev $v$. In the seesaw limit, $M\gg m_D$, the spectrum
of neutrino masses splits into a light set given by the eigenvalues $m_1<m_2<m_3$
of the neutrino mass matrix
\be\label{seesaw}
m_{\nu} = - m_D\,{1\over M}\,m_D^T \, ,
\ee
and into a heavy set $M_1 < M_2 < M_3$ coinciding to a good
approximation with the eigenvalues of the Majorana mass matrix
corresponding to eigenstates $N_i\simeq N_{i R} + N_{i R}^c$.
The symmetric neutrino mass matrix $m_{\nu}$ is diagonalized by a unitary matrix $U$,
\be
D_m \equiv {\rm diag}(m_1,m_2,m_3) = -U_{\nu}^{\dagger}\,m_{\nu}\,U_{\nu}^{\star} \, .
\ee
In a basis where the charged lepton mass matrix is diagonal,
$U_{\nu}$ coincides with the leptonic mixing matrix $U$.
In this way the lightness of ordinary neutrinos is explained just
as an algebraic by-product.
If the largest eigenvalue in the Dirac neutrino mass matrix
is assumed to be of the order of the electroweak scale, as for the other massive fermions,
then the atmospheric neutrino mass scale $m_{\rm atm}$ can be
naturally reproduced for $M_3 \sim 10^{14-15}\,{\rm GeV}$,
very close to the grand-unified scale.  This is the minimal version of the seesaw mechanism,
often indicated as type I seesaw mechanism.

The seesaw formula~(\ref{seesaw}) can be recast as an orthogonality
condition for a matrix $\Omega$ that in a basis where simultaneously the
charged lepton mass matrix and the Majorana mass matrix are diagonal,
provides a useful parametrisation of the neutrino Dirac mass matrix \cite{Casas:2001sr},
\be\label{casas}
m_D=U\,D_m^{1/2}\,\O\, D_M^{1/2} \,  ,
\ee
where $D_M\equiv {\rm diag}(M_1,M_2,M_3)$.  The orthogonal matrix contains 6 
independent parameters and together with the neutrino masses,
determines the properties of the heavy RH neutrinos, the three lifetimes and the
three total $C\!P$ asymmetries.  In this way, one can easily see
that the total number of additional parameters introduced by the see-saw lagrangian is 18 
in the case of 3 RH neutrinos. The orthogonal parametrisation is quite useful since, 
given a model that specifies $m_D$ and the three RH neutrino masses $M_i$,
one can easily impose both the experimental
information on the $9$ low-energy neutrino parameters (6 contained in $U$ and the 3 $m_i$) and, as we will
see, also the requirement of successful leptogenesis.

\section{The simplest scenario: vanilla leptogenesis}

Leptogenesis  belongs to a class of models of Baryogenesis
where the asymmetry is generated from the the out-of-equilibrium
decays of very heavy particles. Interestingly,
this is the same class to which the first model
proposed by Sakharov belongs. This class of models became very popular with the
advent of Grand Unified Theories (GUT) that provided a well-defined and motivated framework.
In GUT baryogenesis models the very heavy decaying
particles generating the asymmetry are the same new gauge bosons predicted by the GUT's.
However,  the final asymmetry depends on too many untestable parameters,
so that imposing successful baryogenesis does not lead to
compelling experimental predictions.
This lack of predictability is made even stronger
considering that the decaying particles are
too heavy to be produced thermally and one has, therefore, to invoke
a non-thermal production mechanism of the gauge bosons.
This is because while the mass of the gauge bosons is
about the GUT scale, $M_X \sim 10^{15-16}\,{\rm GeV}$,
the reheating temperature at the end of inflation $T_{\rm reh}$ cannot be higher
than $\sim 10^{15}\,{\rm GeV}$ from CMB observations, where
the reheating temperature is  the value of the temperature at the beginning of
the radiation-dominated regime after inflation \cite{reheating}.

The minimal (and original) version of leptogenesis, that we discuss in this review, is based on the
type I seesaw mechanism~\cite{Fukugita:1986hr}, where the  asymmetry is produced
by the three heavy RH neutrinos predicted by the see-saw mechanism and whose  masses
can have values orders-of-magnitude below the upper bound on the $T_{\rm reh}\lesssim 10^{15}\,{\rm GeV}$ from
inflation and CMB observations.
 We will call `minimal leptogenesis scenarios' those scenarios
based on a type I seesaw mechanism
and on a thermal production
of the RH neutrinos, implying that $T_{\rm reh}$ cannot be too much below (at least)
the lightest RH neutrino mass $M_1$
\footnote[3]{For a description of leptogenesis from RH neutrino oscillations, and for 
leptogenesis scenarios based on models
beyond the type I see-saw mechanism, see respectively \cite{canetti} and  \cite{hambye} in this Focus Issue.}. 
At these high temperatures the RH neutrinos can be produced
by the Yukawa interactions of leptons and Higgs bosons in
the thermal bath. A solution of the Boltzmann equations shows that, more
exactly, $T_{\rm reh}$ can be even up to $10$ times lower than $M_1$,
the exact value depending on the strength of the coupling \cite{pedestrians}.

After their production, the RH neutrinos decay either
into lepton and Higgs doublets $N_i \ra {\ell}_i + \Phi$ (when leptogenesis
takes place, the electroweak symmetry is not broken) with a decay rate $\Gamma$,
or into anti-lepton and conjugate Higgs doublets $N_i \rightarrow \bar{\ell}_i + \Phi^{\dagger}$
with a decay rate $\bar{\Gamma}$.
Both inverse processes and decays violate lepton number ($\Delta L=1$)
and in general $C\!P$ as well.
It is also important to notice that they also violate the $B-L$
asymmetry, ($\D(B-L)=1$).
At temperatures $T\gg 100\,{\rm GeV}$,  non-perturbative Standard Model processes called
sphalerons are in equilibrium \cite{Kuzmin:1985mm}. They violate
both lepton and baryon number while they still conserve $B-L$.
Hence the lepton asymmetry produced in the elementary processes is
reprocessed in a way that $\sim 1/3$ of the $B-L$ asymmetry
is in the form of baryon number  while  $\sim -2/3$ of the $B-L$ asymmetry
is in the form of a lepton number \cite{harvey}. Therefore, two of the Sakharov conditions
are satisfied. The third Sakharov condition, departure from thermal equilibrium,
is also satisfied since a fraction of the decays occur out of equilibrium, implying that
part of the asymmetry survives the washout from inverse processes.

In the most general approach, even within minimal
leptogenesis, the asymmetry depends on all the 18 seesaw parameters
and the calculation itself presents different technical difficulties.
However, there is  a simplified  scenario
\cite{Fukugita:1986hr,Luty:1992un,bcst,cmb,window,giudice,pedestrians,bounds} that grasps
most of the main features of
leptogenesis and is able to highlight  important connections
with the low-energy neutrino parameters in an approximated way.
We will refer to this  scenario as `vanilla leptogenesis'.
Let us  discuss  the main assumptions and approximations.

The first assumption is the one-flavour approximation, where
the flavour composition of the leptons
produced by (or producing) the RH neutrinos has no influence on the
final value of the asymmetry and can be therefore neglected.  It also assumes
that leptons produced by different RH neutrinos can be treated, for all practical
purposes, as having the same flavour (i.e. ${\ell}_1 = {\ell}_2 = {\ell}_3 = {\ell}$).
This is equivalent to saying that what counts is just the asymmetry between the
total number of leptons and the total number of anti-leptons irrespectively of
whether these leptons are electron, muon or tauon doublets, and irrespectively of the
RH neutrino that generated it. Under this assumption the
final $B-L$ asymmetry is the sum of two contributions
\be\label{NBmLf}
N^{\rm f}_{B-L} = N^{\rm pre-ex,f}_{B-L}+ N^{\rm lep,f}_{B-L}\, .
\ee
The first term on the right-hand side, $N^{\rm pre-ex,f}_{B-L}$, is the residual value of a possible
pre-existing asymmetry generated by some external mechanism prior to
the onset of leptogenesis. It has to be regarded as a possible external contribution whose initial
value would be set up by some  independent source of leptogenesis. Therefore, if leptogenesis
is responsible for the observed asymmetry, this term has to be negligible.

The second term is the genuine leptogenesis contribution to the final asymmetry and
is given by the sum of the three contributions from each RH neutrino species,
\be
N^{\rm lep,f}_{B-L}= \sum_i \, \ve_i \, \kappa_i^{\rm f}(K_1, K_2, K_3) \, .
\ee
Each contribution is the product of the total $C\!P$ asymmetry $\ve_i$
times  the final value of the efficiency factor, $\k_i^{\rm f}(K_1, K_2, K_3)$,
which depends on  the decay parameters $K_i$, defined as $K_i\equiv \Gamma_{{\rm D}, i}(T=0)/H(T=M_i)$,
where $H$ is the Hubble expansion rate, and the total decay rates are
$\G_{{\rm D},i} \equiv \G_i+\bar{\G}_i$.
 The total $C\!P$  asymmetries are defined  as
\be
\ve_i \equiv - {\G_i-\overline{\G}_i \over \G_i+\overline{\G}_i} \, ,
\ee
and a perturbative calculation from the
interference of tree level with one
loop self-energy and vertex diagrams gives \cite{flanz,Covi:1996wh,buchplumi1}
\be\label{CPas}
\ve_i =\, {3\over 16\pi}\, \sum_{j\neq i}\,{{\rm
Im}\,\left[(h^{\dagger}\,h)^2_{ij}\right] \over
(h^{\dagger}\,h)_{ii}} \,{\xi(x_j/x_i)\over \sqrt{x_j/x_i}}\, ,
\ee
having introduced
\be\label{xi}
\xi(x)= {2\over 3}\,x\,
\left[(1+x)\,\ln\left({1+x\over x}\right)-{2-x\over 1-x}\right] \, ,
\ee
and defined $x_j \equiv M_j^2/M_1^2$. The efficiency factors
are computed solving a simple set of Boltzmann equations integrated
over the momenta (rate equations)
\begin{eqnarray}
{dN_{N_i}\over dz} & = &
-D_i\,(N_{N_i}-N_{N_i}^{\rm eq}) \;,
\hspace{10mm} i=1,2,3 \label{dlg1} \\\label{unflke}
{dN_{B-L}\over dz} & = &
\sum_{i=1}^3\,\varepsilon_i\,D_i\,(N_{N_i}-N_{N_i}^{\rm eq})-
N_{B-L}\,\left[\D W(z)+\sum_i \,W_i^{\rm ID}(z)\right] \;  ,
\end{eqnarray}
where $z \equiv M_1/T$ and where we indicated with $N_X$
any particle number or asymmetry $X$ calculated in a portion of co-moving
volume containing one heavy neutrino in ultra-relativistic thermal equilibrium,
so that $N^{\rm eq}_{N_i}(T\gg M_i)=1$.
With this convention, the predicted baryon-to-photon ratio $\eta_B$ is
related to the final value of the final $B-L$ asymmetry by the relation
\be\label{etaB}
\eta_B=a_{\rm sph} {N_{B-L}^{\rm f}\over N_{\g}^{\rm rec}}\simeq 0.96\times
10^{-2} N_{B-L}^{\rm f}\, ,
\ee
where $N_{\g}^{\rm rec}\simeq 37$, and $a_{\rm sph}=28/79$.
Defining $z_i\equiv z\,\sqrt{x_i}$,
the decay factors are defined as
\be
D_i \equiv {\G_{{\rm D},i}\over H\,z}=K_i\,x_i\,z\,
\left\langle {1\over\gamma_i} \right\rangle   \, ,
\ee
where the
thermally averaged dilation factors $\langle 1/\gamma\rangle$ are given by the ratio
${\cal K}_1(z)/ {\cal K}_2(z)$ of the modified Bessel functions.
After proper subtraction of the resonant contribution from
$\Delta L=2$ processes in order to avoid double counting~\cite{Kolb:1979qa,giudice}, the inverse decay
washout terms are simply given by
\be\label{WID}
W_i^{\rm ID}(z) =
{1\over 4}\,K_i\,\sqrt{x_i}\,{\cal K}_1(z_i)\,z_i^3 \, .
\ee
The washout term $\D W (z)$ is the non-resonant $\D L=2$ processes contribution.
It gives a non-negligible effect only at $z\gg 1$ and in this case
it can be approximated as \cite{cmb}
\be
\Delta W(z) \simeq {\o \over z^2}\,\left(M_1\over 10^{10}\,{\rm GeV}\right)\,
\left({\overline{m}^{\, 2} \over {\rm eV^2}}\right) \, ,
\ee
where $\o\simeq 0.186$ and $\overline{m}^2\equiv m_1^2+m_2^2+m_3^2$.
The efficiency factors can be simply calculated analytically as
\be\label{efial}
\k_{i}^{\rm f}(K_i)=-\int_{z_{\rm in}}^\infty\,dz'\,{dN_{N_i}\over dz'}\,
{\rm e}^{-\int_{z'}^\infty\,dz''\,[\D W(z'')+\sum_i\,W_i^{\rm ID}(z'';K_i)]} \,.
\ee

 The second assumption is that the RH neutrino mass spectrum is hierarchical
with  $M_2 \gtrsim 3\, M_1$.

Under these two assumptions (i.e. unflavoured description plus hierarchical RH neutrino mass spectrum)
and barring a particular case, that we will discuss later on,
the final asymmetry  is typically dominantly produced by the lightest RH neutrino out-of-equilibrium decays,
in  a way that the sum in Eq.~(\ref{NBmLf}) can be approximated by
the first term ($i=1$), explicitly
\be\label{unflavoured}
N_{B-L}^{\rm f}\simeq \ve_1\,\k_1^ {\rm f}(K_1)    \,  ,
\ee
so that an `$N_1$-dominated scenario' is realised.
This happens  either because $|\ve_{2,3}|\ll |\ve_1|$ or because
the asymmetry initially produced
by the $N_{2,3}$ decays is afterwards washed out
by the lightest RH neutrino inverse processes such that
$\k_{2,3}^{\rm f}(K_{2,3})\ll \k_1^{\rm f}(K_1)$.  Indeed,
if we indicate with $N_{B-L}^{(2,3)}(T\gtrsim M_1)$ the contribution
to the $N_{B-L}$ asymmetry from the two heavier RH neutrinos prior
to the lightest RH neutrino washout, the final values are given simply by
\be
N_{B-L}^{(2,3), {\rm f}} = N_{B-L}^{(2,3)}(T\gtrsim M_1) \, e^{-{3\pi \over 8}\,K_1} \, .
\ee
The same exponential washout factor also suppresses the residual value
of a possible pre-existing asymmetry. Therefore, it is sufficient
to impose the strong washout condition $K_1 \gg 1$
for the pre-existing asymmetry
and for the contribution from heavier RH neutrinos to be negligible.
The decay parameters  can be expressed in terms of the see-saw parameters as
\be
K_i={(h^{\dagger}h)_{ii} v^2\over M_i \, m_{\star}} \, ,
\ee
where $m_{\star}\simeq 1.1\times 10^{-3}$~{\rm eV} is the equilibrium neutrino mass \cite{bcst,window}.
It is therefore quite straightforward that,
barring cancellations in the seesaw formula,
one typically has $ K_i \sim K_{\rm sol}$--$K_{\rm atm} \simeq 10 $--$ 50\gg 1$, where
$K_{\rm sol(atm)}\equiv m_{\rm sol(atm)}/m_{\star}$.
The same condition  also guarantees that the final asymmetry is independent of the initial
$N_1$-abundance. It is then quite suggestive that
the measured values of $m_{\rm sol}$ and $m_{\rm atm}$ have just the right values
to produce a washout that is strong enough to guarantee independence of the initial
conditions but still not that strong to prevent successful leptogenesis \cite{window}. This {\em leptogenesis
conspiracy} between experimental results and theoretical predictions is the main reason
that has determined the success of leptogenesis during the last years.

There is a particular case where $K_1 \gg 1$ does not hold and in this case
the final asymmetry can be dominated by the contribution coming from
the next-to-lightest  RH neutrinos \cite{DiBari:2005st}.
For this case one still has $K_2 \gg 1$, so that
the independence of the initial conditions holds nonetheless.
For the time being, as an additional
third assumption, we will not consider this particular case.

If one excludes cancellations among
the different terms contributing to the neutrino masses
in the seesaw formula, one obtains the Davidson-Ibarra upper bound
on the lightest RH neutrino $C\!P$ asymmetry \cite{Davidson:2002qv}
\be\label{CPbound}
\ve_1 \leq \ve_1^{\rm max}
\simeq 10^{-6}\,{M_1\over 10^{10}\,{\rm GeV}}\,{m_{\rm atm}\over m_1+m_3} \, .
\ee
Imposing $\eta_B^{\rm max}\simeq 0.01\,\ve_1^{\rm max}\,\k_1^{\rm f} > \eta_{B}^{\rm CMB}$,
one obtains the allowed region in the plane $(m_1,M_1)$ shown in the left panel of Fig.~2.
\begin{figure}
\begin{center}
\psfig{file=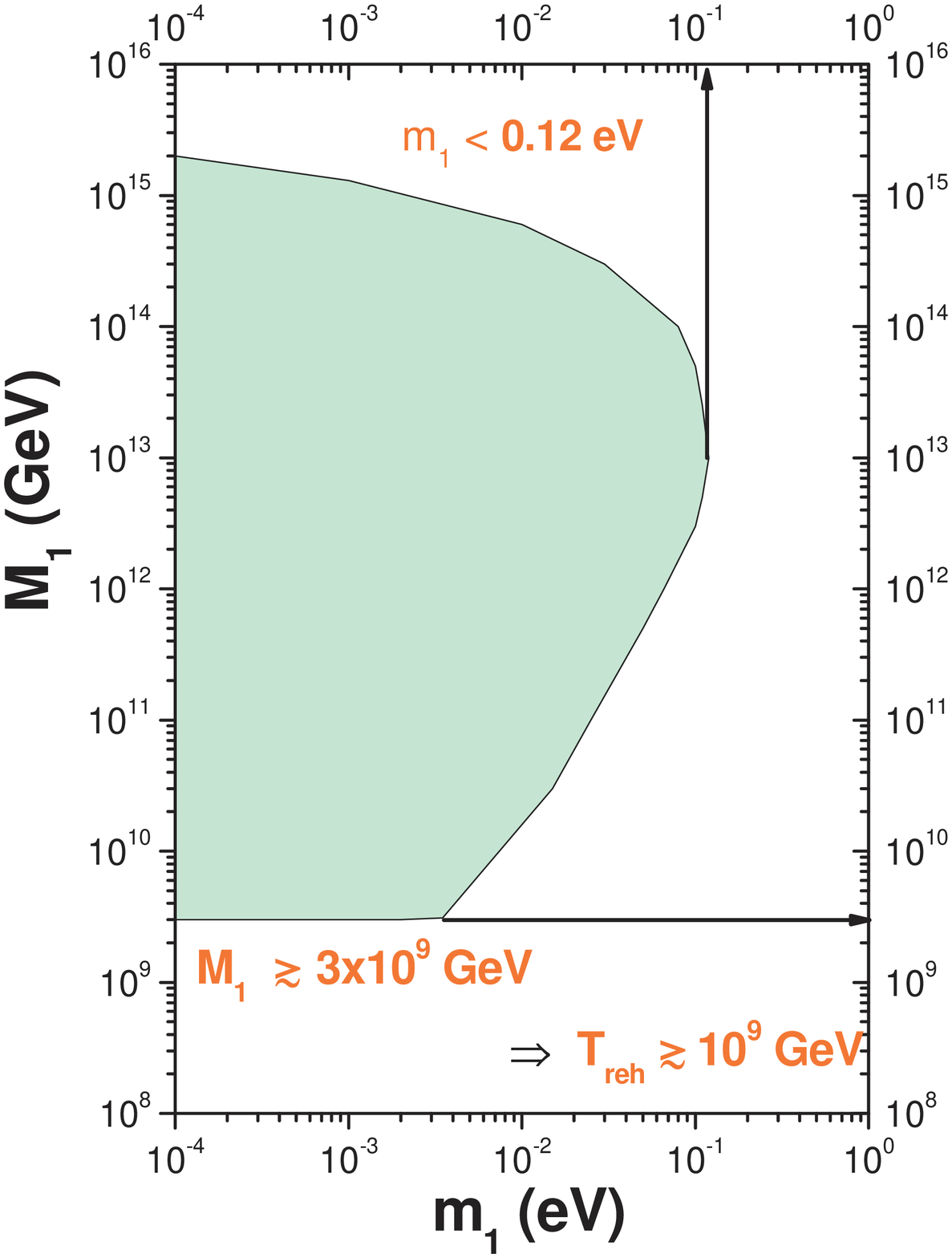,width=0.4\textwidth}  \hspace{5mm}
\psfig{file=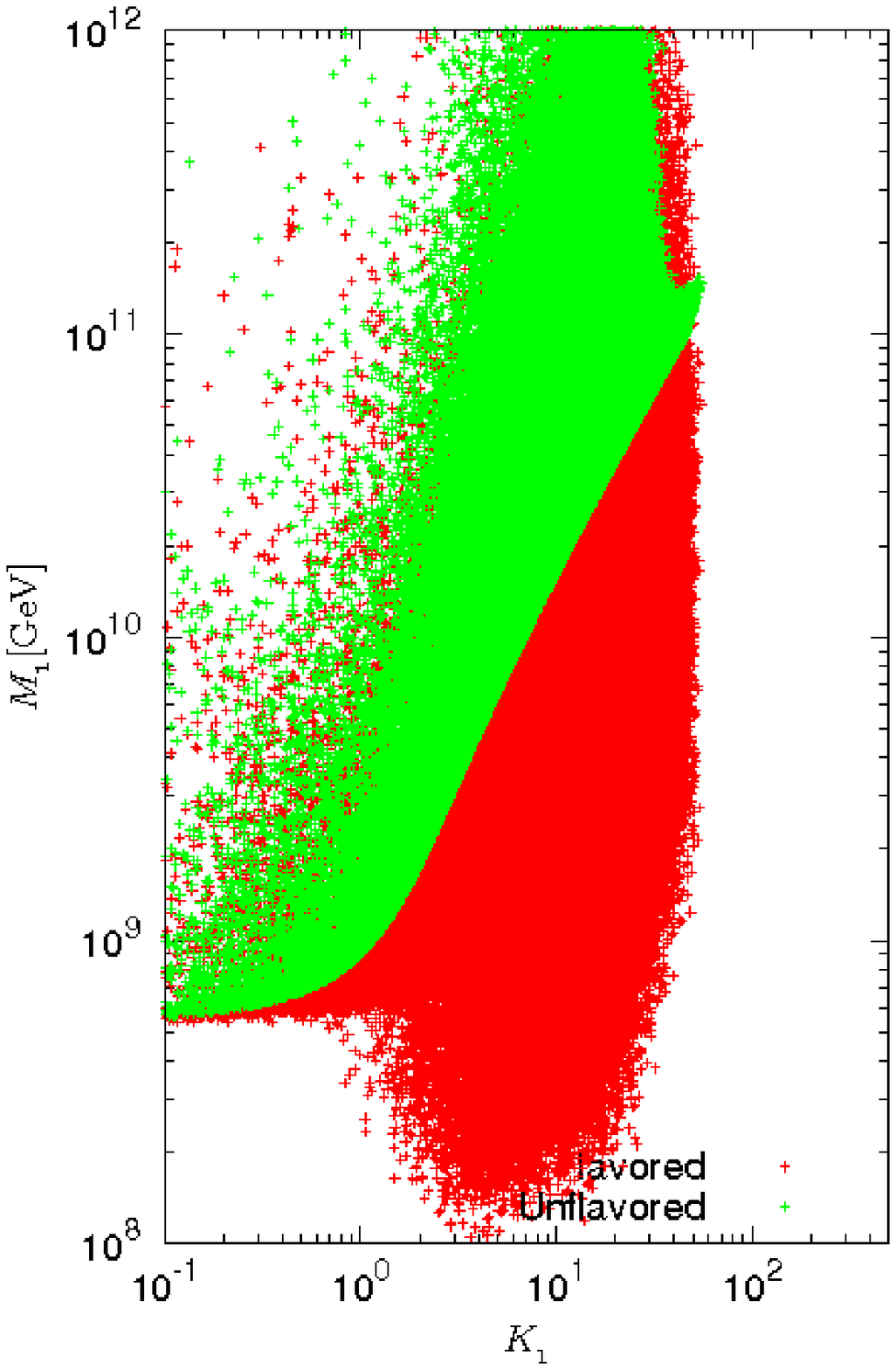,width=0.4\textwidth}
\end{center}
\caption{Left: Neutrino mass bounds in the vanilla scenario. Right:
         Relaxation of the lower bound on $M_1$ thanks
          to additional unbounded flavoured $C\!P$ violating terms~\cite{bounds}.}
\end{figure}
One can notice the existence of an upper bound on the light
neutrino masses $m_1\lesssim 0.12\,{\rm eV}$ \cite{Buchmuller:2002jk,window},
incompatible with quasi-degenerate
neutrino mass models, and a lower bound on
$M_1\gtrsim 3\times 10^9\,{\rm GeV}$ \cite{Davidson:2002qv,Buchmuller:2002jk,window}
implying a lower bound on the
reheat temperature $T_{\rm reh}\gtrsim 10^9\,{\rm GeV}$ \cite{pedestrians}.

An important feature of vanilla leptogenesis is that the final asymmetry does not
directly depend on the parameters of leptonic mixing matrix $U$.
That implies that one cannot establish a direct model-independent connection.
In particular the discovery
of $C\!P$ violation in neutrino mixing would not be a smoking gun for leptogenesis
and, vice versa, a non-discovery would not rule out leptogenesis.
However, within more restricted scenarios, for example imposing some conditions on
the neutrino Dirac mass matrix, links can emerge. In Section 9 we will discuss
in detail the interesting case of $SO(10)$-inspired models.

In  the past years different directions beyond the
vanilla leptogenesis scenario were explored,
usually aiming at evading the above-mentioned bounds.
For example, it was noticed that the Davidson-Ibarra bound eq.~(\ref{CPbound})
is strictly speaking evaded by an extra-term contribution $\Delta \ve_1$ to the total
$C\!P$ asymmetry \cite{hambyestrumia,bounds}. However, this extra-term vanishes
if $M_3 \simeq M_2$ and is suppressed as $\Delta \ve_1 \propto (M_1/M_2)^2$. It can
significantly relax the bounds on neutrino masses only when $|\Omega_{ij}|^2 \gtrsim (M_2/M_1)^2$,
implying cancellations with a certain degree of fine tuning in the see-saw formula 
for the neutrino masses. For example,
in usual models with $|\Omega_{ij}|^2 \lesssim 10$, this extra-contribution can be safely neglected
for $M_2\gtrsim 10\,M_1$.

The development that proved to have the most important
impact on the final asymmetry, compared to a calculation within the vanilla scenario,
is certainly the inclusion of flavour effects. For this reason we discuss
them in detail in the next Section.

\section{The importance of flavour}

The inclusion of flavour effects provides the most significant modification in
the calculation of the final asymmetry compared to the vanilla scenario. 
Two kinds of flavour
effects are neglected in the vanilla scenario:
heavy neutrino flavour effects, how heavier RH neutrinos
can contribute  to the final asymmetry, 
and lepton flavour effects,
how the flavour composition of the leptons quantum states produced in the RH neutrino decays
affects the calculation of the final asymmetry.
We first discuss the two effects separately, and then we show
how their interplay can have very interesting consequences.

\subsection{Heavy neutrino flavour effects}

In the vanilla scenario the contribution to the final asymmetry from the
heavier RH neutrinos is negligible because either the $C\!P$ asymmetries  are
suppressed in the hierarchical limit compared to
$\ve_1^{\rm max}$ (cf. Eq.~(\ref{CPbound})) and/or because,
even assuming that a sizeable asymmetry (compared to the observed one) 
is produced at $T \sim M_{2,3}$,
it is later on washed out by the lightest RH neutrino inverse processes.

However, as we anticipated, there is a particular case  when,
even neglecting the lepton flavour composition and assuming
a hierarchical heavy neutrino mass spectrum,  the contribution
to the final asymmetry from next-to-lightest RH neutrino decays can be
dominant and explain the observed asymmetry \cite{DiBari:2005st}.  This case corresponds to a particular
choice of the orthogonal matrix such that $N_1$ is so weakly coupled
(corresponding to $K_1 \ll 1$) that its washout can be neglected.
For the same choice of the parameters, the $N_2$ total $C\!P$ asymmetry
$\ve_2$ is unsuppressed if $M_3\lesssim 10^{15}\,{\rm GeV}$.
In this case a $N_2$-dominated scenario is realised.
Notice that in this case the existence of a third (heavier) RH neutrino
species is crucial in order to have a sizeable $\ve_2$.

The contribution from the two heavier RH neutrino species  is also important
in the quasi-degenerate limit when $\d_{i}\equiv (M_i-M_1)/M_1 \ll 1,~i=2,3$.
In this case the $C\!P$ asymmetries $\ve_{2,3}$ are not suppressed, and the
washout from the lighter RH neutrino species is moderate, with no exponential
prefactor \cite{Pilaftsis:1997jf,Blanchet:2006dq}.

\subsection{Lepton flavour effects}

The importance and generality of flavour effects in leptogenesis was fully highlighted in \cite{Abada:2006fw,Nardi:2006fx}.
Their  role  was first discussed in \cite{bcst} and 
included in specific scenarios in \cite{endoh,Pilaftsis:2005rv,vives}.

For the time being, let us continue to assume that the final asymmetry is
dominantly produced from the decays of the lightest RH neutrinos $N_1$,
neglecting the contribution from the decays of the heavier RH neutrinos $N_2$ and $N_3$.
If $M_1\gg 10^{12}\,{\rm GeV}$, the flavour composition
of the quantum states of the leptons produced from $N_1$ decays
has no influence on the final asymmetry and a one-flavour
regime holds \cite{bcst,Abada:2006fw,Nardi:2006fx}.
This is because the lepton quantum states evolve coherently  between the production
from a $N_1$-decay
and a subsequent inverse decay with a Higgs boson.
In this way the lepton flavour composition does not play any role.

However, if $10^{12}\,{\rm GeV}\gtrsim M_1 \gtrsim 10^{9}\,{\rm GeV}$,
during the relevant period of generation of the asymmetry, the produced lepton
quantum states will, on average, have an interaction with RH tauons before
undergoing the subsequent inverse decay. In this way the tauon component of the lepton quantum
states is measured by the thermal bath and the coherent evolution breaks down
\cite{bcst,Abada:2006fw,Nardi:2006fx}.
Therefore, at the subsequent inverse decays,
the lepton quantum states are an incoherent mixture of a tauon component and
of a (still coherent) superposition of an electron and  a muon component that
we can indicate with $\tau^{\bot}$.

The fraction of asymmetry stored in each flavour component is not proportional in general
to the branching ratio of that component. This implies that the two
flavour asymmetries, the tauon and the $\tau^{\bot}$ components,
evolve differently and have to be calculated separately.
In this way  the resulting final asymmetry can considerably differ
from the result in the one-flavour regime.  This can be indeed approximated by 
the expression
\be\label{twofully}
N^{\rm f}_{B-L} \simeq 2\,\ve_1\,\k(K_1) +
{\D p_{1\tau}\over 2}\,\left[\k(K_{1{\tau}^{\bot}_1})-\k(K_{1\tau})\right] \, ,
\ee
where $K_{1\a}\equiv p^0_{1\a}\,K_i$,  the $p^0_{1\a}$'s ($\a=\t, \t^{\bot}$) are the 
tree level probabilities that the leptons ${\ell}_1$ and the anti-leptons $\bar{\ell}'_1$ 
produced in the decays of the $N_1$'s 
are in the flavour $\a$, while $\D p_{1\tau}$ is the difference 
between the probability to find ${\ell}_1$ in the flavour $\tau$ and that
one to find $\bar{\ell}'_1$ in the flavour $\bar{\tau}$.  If we 
compare this expression with the one-flavour regime result eq.~(\ref{unflavoured})
one can see that if $\D p_{1\tau}=0$ (or if $K_{1\tau}=K_{1\tau^{\bot}}$), 
then the final symmetry is enhanced just by a factor $2$. However, in general
leptons and anti-leptons have a different flavour composition \cite{bcst,Nardi:2006fx}
and in this case $\D p_{1\tau}\neq 0$ and the final asymmetry can be much higher than in the one-flavour regime.
The most extreme case is when the total $B-L$ number is conserved (i.e. $\ve_1=0$)
and still the second term in eq.~(\ref{twofully}) 
can be non-vanishing \cite{Nardi:2006fx} and even explain the observed asymmetry \cite{Blanchet:2006be}. 

If $M_1\lesssim 10^{9}\,{\rm GeV}$, even the coherence of the $\tau^{\bot}$
component is broken by the muon interactions between decays and inverse decays
and a full three flavour regime applies. In the intermediate regimes
a density matrix formalism is necessary to properly describe  decoherence
\cite{Abada:2006fw,Blanchet:2006ch,DeSimone:2006dd,Beneke:2010dz}.

We can briefly say that 
lepton flavour effects induce three major consequences that are all encoded 
in the expression (\ref{twofully}) valid in the two fully flavoured regime. 
i) The washout can be considerably lower than in the unflavoured regime \cite{Abada:2006fw,Nardi:2006fx}
since one can have that the asymmetry is dominantly produced in a flavour $\a$ with $K_{1\a}\ll K_1$
ii) The leptonic mixing matrix enters directly in the calculation of the final asymmetry, more specifically
and in particular the low-energy phases
contribute as a second source of $C\!P$ violation in the flavoured $C\!P$ asymmetries
\cite{Nardi:2006fx,Abada:2006ea,Blanchet:2006be,Pascoli:2006ci}. As an interesting phenomenological 
consequence, the same source of $C\!P$
violation that could take place in neutrino oscillations could be sufficient
to explain the observed  asymmetry \cite{Blanchet:2006be,Pascoli:2006ci},
though under quite stringent conditions on the RH neutrino mass spectrum 
and with an exact determination requiring a density matrix calculation \cite{Anisimov:2007mw}.
Notice that this is a particular case  realising the above-mentioned scenario with $\ve_1=0$ \cite{Nardi:2006fx}.   
This problem becomes particularly interesting
in the light of the recent discovery of a non-vanishing $\theta_{13}$ angle \cite{hints,T2K,dayabay,reno},
a necessary condition to have $C\!P$ violation in neutrino oscillations.
 In particular a calculation of the lower bound on $\theta_{13}$ that necessarily requires
the use of density matrix equations.
iii) The flavoured $C\!P$ asymmetries, given by
\be\label{flavouredCP} \fl
\varepsilon_{i\alpha}=
{3\over 16\pi (h^{\dagger}h)_{ii}}\sum_{j\neq i} \left\{ {\rm Im}
\left[h_{\alpha i}^{\star}h_{\alpha j}(h^{\dagger}h)_{ij}\right] {\xi(x_j/x_i)\over
\sqrt{x_j/x_i}}\right. \nonumber
\left.+{2\over 3(x_j/x_i-1)}{\rm Im}
\left[h_{\alpha i}^{\star}h_{\alpha j}(h^{\dagger}h)_{ji}\right]\right\} \, ,
\ee
contain a second term that conserves the total lepton number (it cancels exactly
when summing over flavour in the total $C\!P$ asymmetry and it, therefore, contribute to
$\D p_{1\tau}$ in eq.~(\ref{twofully}) but not to $\ve_1$),
and therefore the upper bound in Eq.~(\ref{CPbound})
does not strictly apply to the flavoured $C\!P$ asymmetries.
As a consequence, allowing for a mild cancelation in the neutrino mass matrix,
corresponding to $|\O_{ij}|\sim 1$, and also
for a mild RH neutrino mass hierarchy ($M_2/M_1 \sim 10$), the lower bound on $T_{\rm reh}$
can be relaxed by about one order of magnitude, down to $10^8\,{\rm GeV}$
\cite{bounds}, as shown in the right panel of Fig.~2.
However, for many models such as sequential dominated models \cite{King:2003jb},
these cancellations do not occur, and lepton flavour effects cannot relax the
lower bound on $T_{\rm reh}$ \cite{Blanchet:2006be}. One known exception is given
by the inverse seesaw model~\cite{Mohapatra:1986bd} which naturally explains large
cancellations in the neutrino
mass matrix and leads to large $C\!P$ asymmetries thanks to an underlying
lepton number symmetry. This leads to the relaxation of the lower bound on $T_{\rm reh}$ 
by up to three orders of magnitude~\cite{Antusch:2009gn,Racker:2012vw}
(see more detailed discussion in the next Section).

Whether or not the upper bound $m_i\lesssim 0.1$~eV on neutrino masses found in the vanilla scenario
still holds in a flavoured $N_1$-dominated scenario, is relaxed, or even completely evaporates,  is a controversial topic. 
It still surely holds in the one-flavour regime for $M_1 \gtrsim 10^{12}\,{\rm GeV}$ but it certainly does not hold in the two-fully flavoured regime
\cite{Abada:2006fw,JosseMichaux:2007zj,bounds}. 
However, it was found that the two-fully 
flavoured regime is not respected at large values  $m_1\gtrsim 0.1 \,{\rm eV}$
\cite{Blanchet:2006ch}.
In \cite{DeSimone:2006dd} it was found 
that it holds up to $m_1 \sim 2\,{\rm eV}$, 
implying in any case an upper bound much above current
experimental bounds and, therefore, uninteresting. 
In \cite{bounds}, including  the information from low energy neutrino data  and accounting for the Higgs asymmetry, it was found again that it 
holds only up to  $m_1\gtrsim 0.1 \,{\rm eV}$ and in this case a density matrix approach  would be required  for a conclusion.

\subsection{The interplay between lepton and heavy neutrino flavour effects}

As we have seen, when lepton flavour effects are neglected, the possibility
that the next-to-lightest RH neutrino decays contribute to the final
asymmetry relies on a special case realising the $N_2$-dominated scenario
\cite{DiBari:2005st}. On the other hand,
when lepton flavour effects are taken into account,
the contribution from heavier RH neutrinos
cannot be neglected in a much more general situation.  Even the contribution
from the heaviest RH neutrinos can be sizeable (i.e. explain the observed asymmetry)
and has to be taken into account in general.

As a result, the calculation of the final asymmetry becomes much more involved.
Assuming hierarchical mass patterns and that the RH neutrino processes
occur in one of the three different fully flavoured regimes, one has to consider
ten different mass patterns, shown in Fig.~2, that require specific multi-stage
sets of classical Boltzmann equations for the calculation of the final asymmetry.
\begin{figure}
\begin{minipage}{140mm}
\begin{center}
\psfig{file=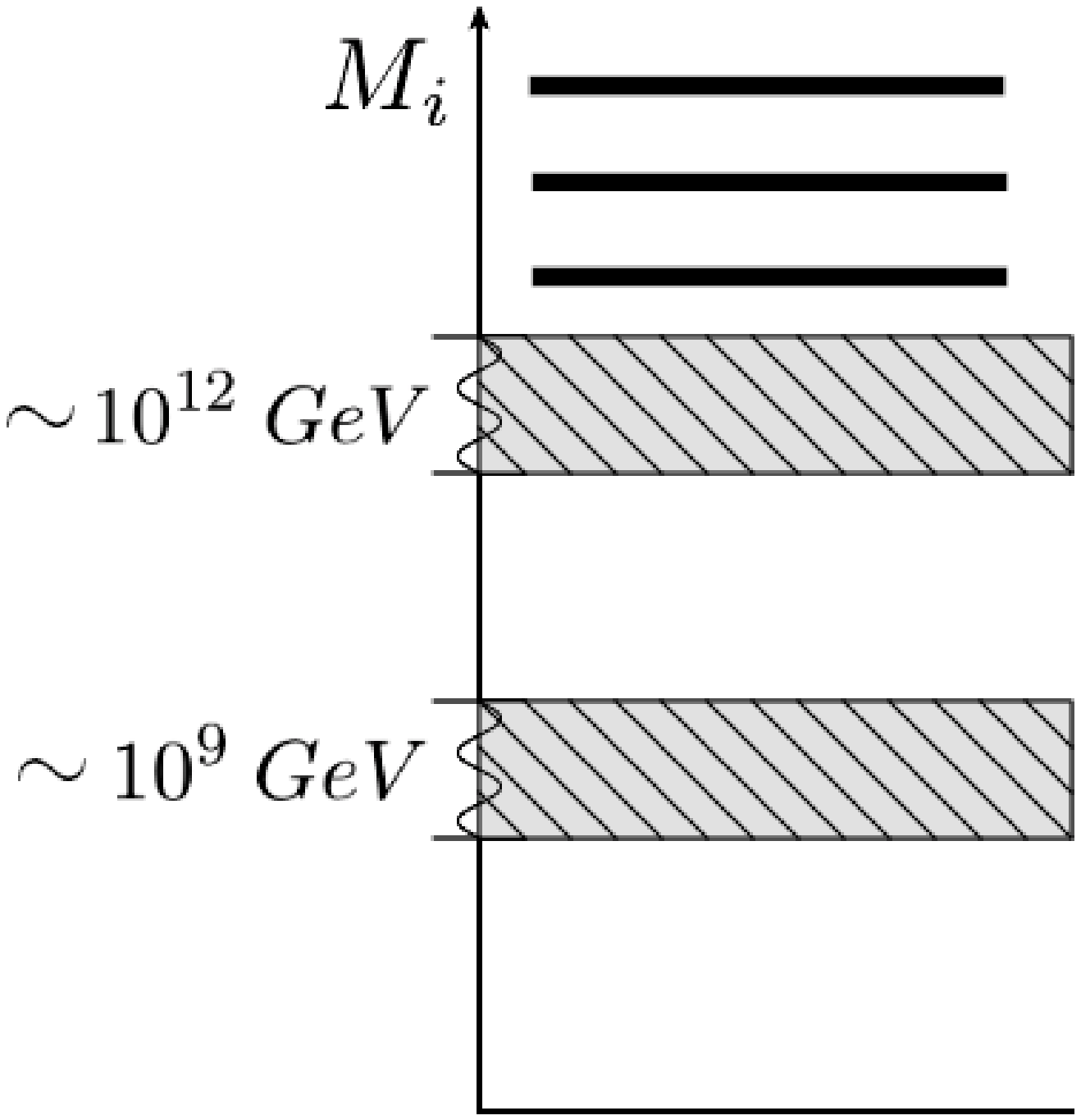,height=3cm,width=3cm} \hspace*{2mm}
\psfig{file=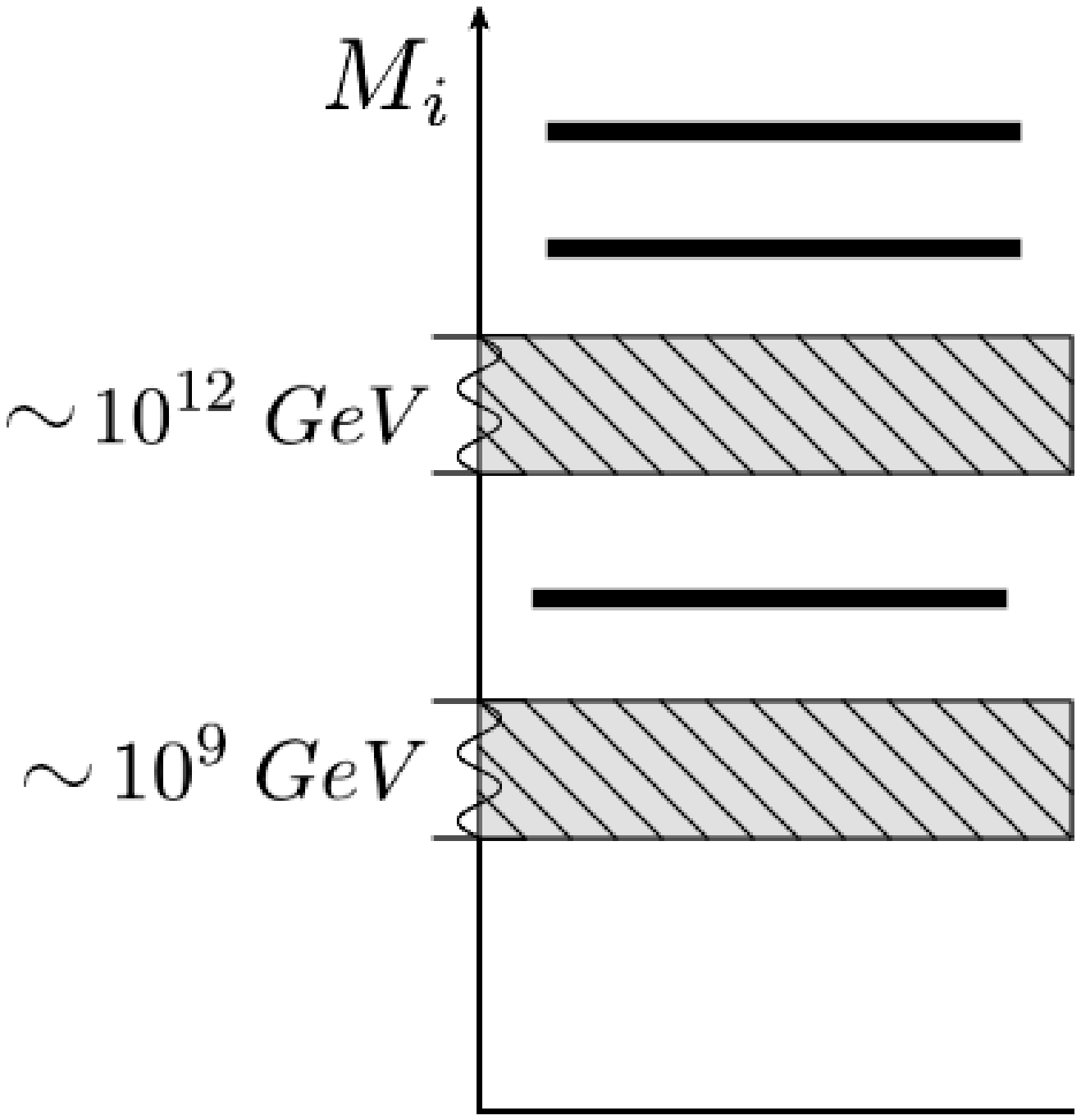,height=3cm,width=3cm} \hspace*{2mm}
\psfig{file=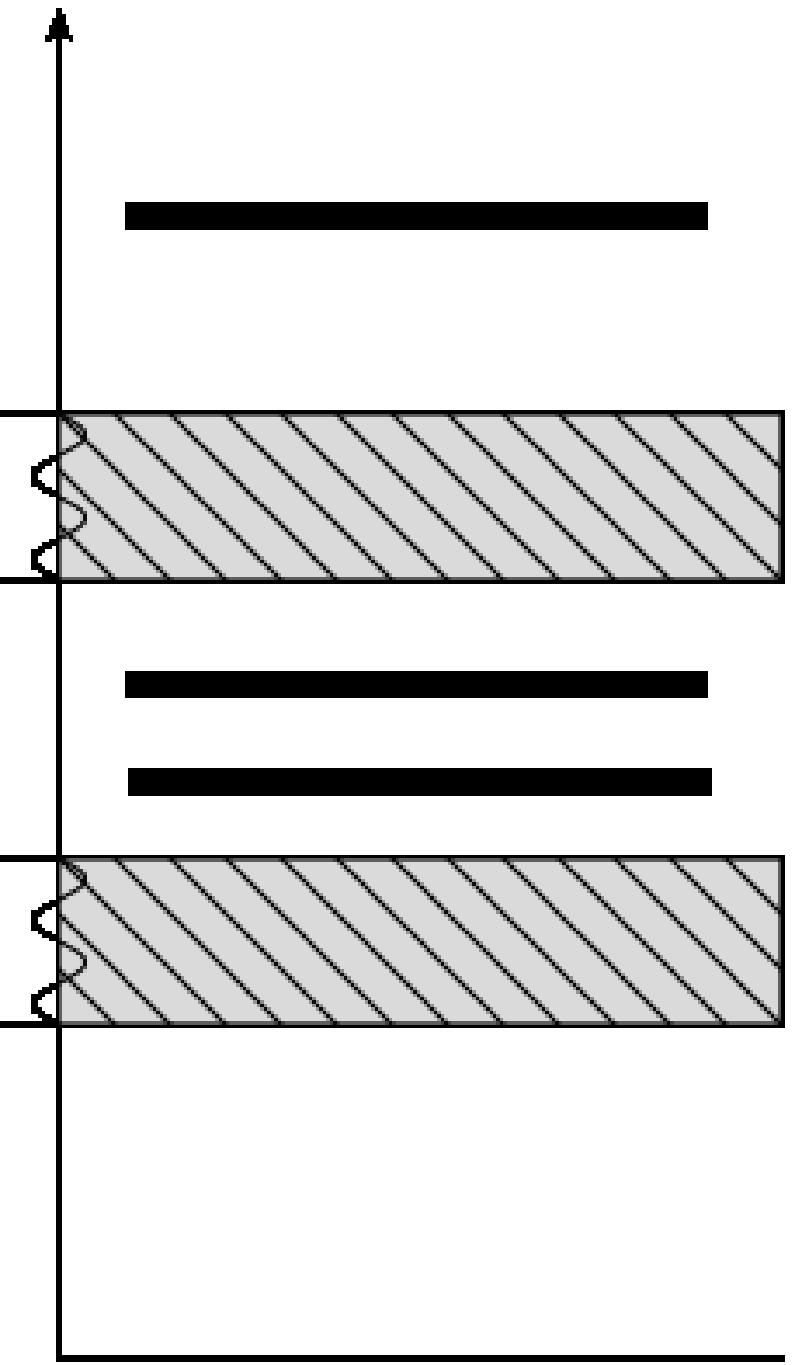,height=3cm,width=15mm} \hspace*{2mm}
\psfig{file=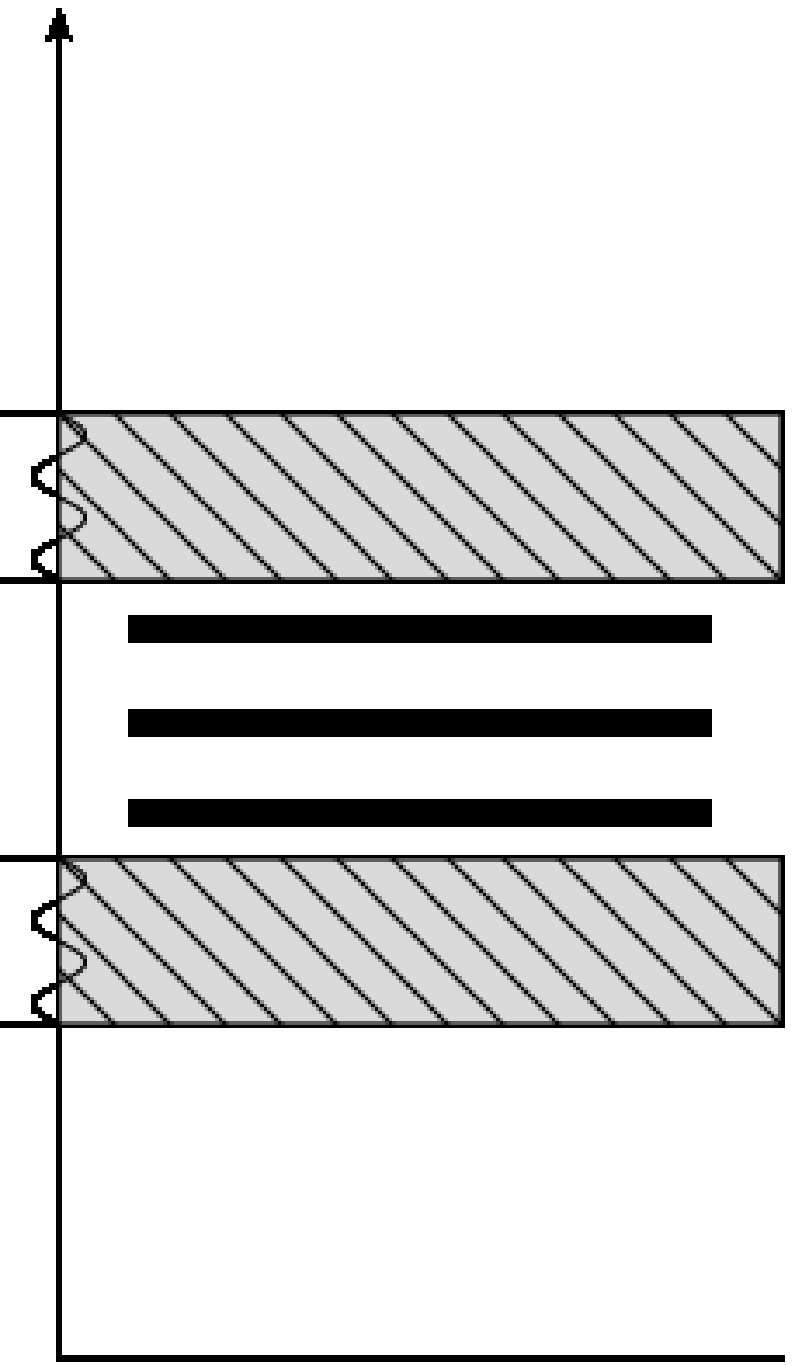,height=3cm,width=15mm}
\end{center}
\begin{center}
\psfig{file=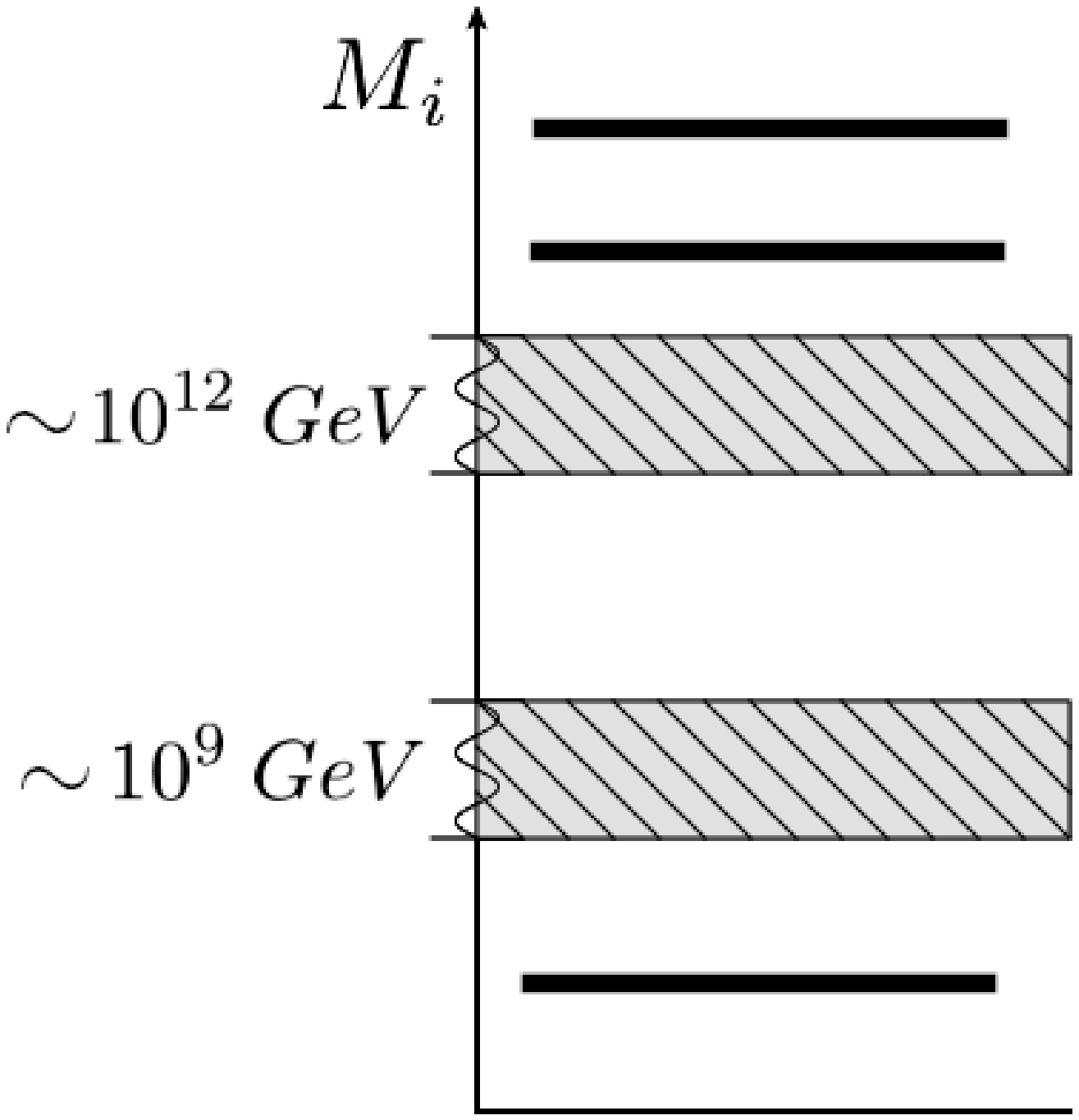,height=3cm,width=3cm}
\psfig{file=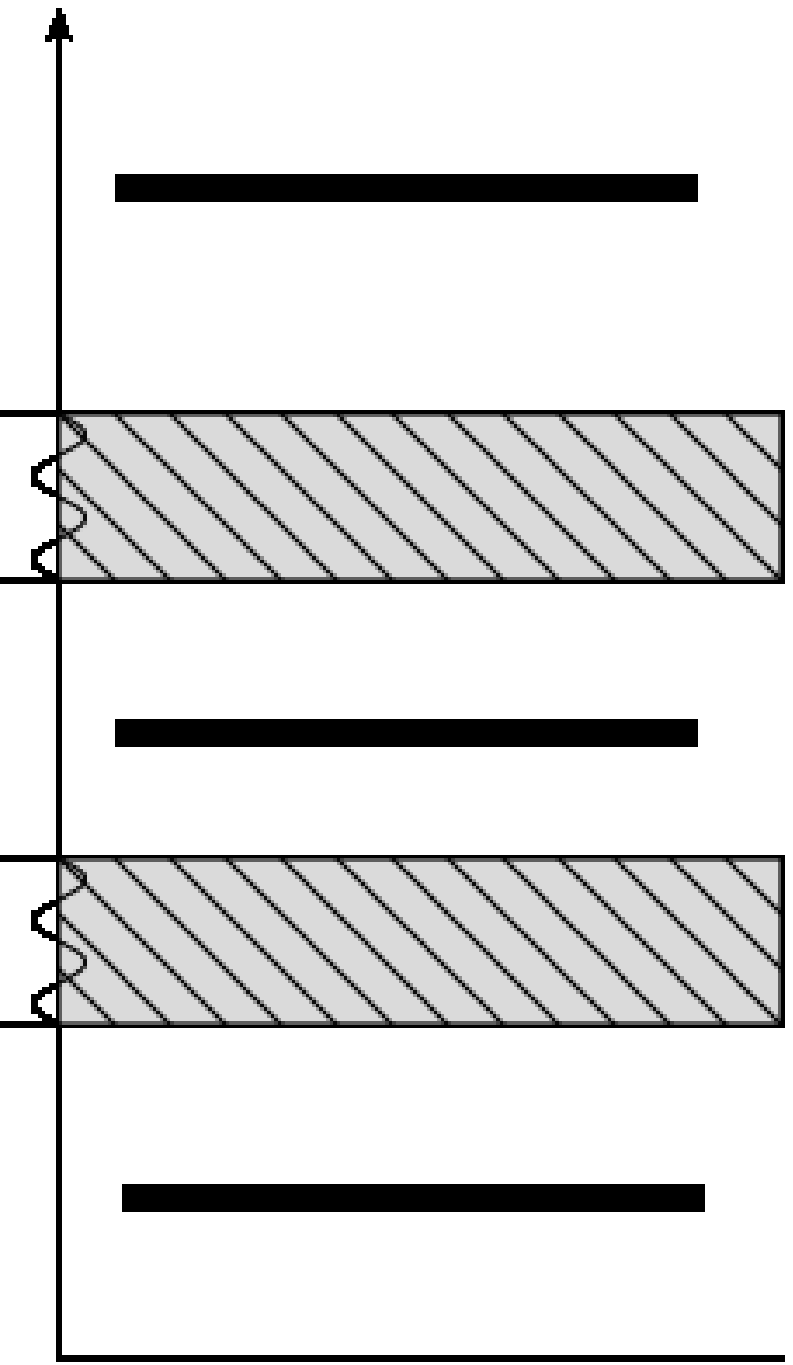,height=3cm,width=15mm}
\psfig{file=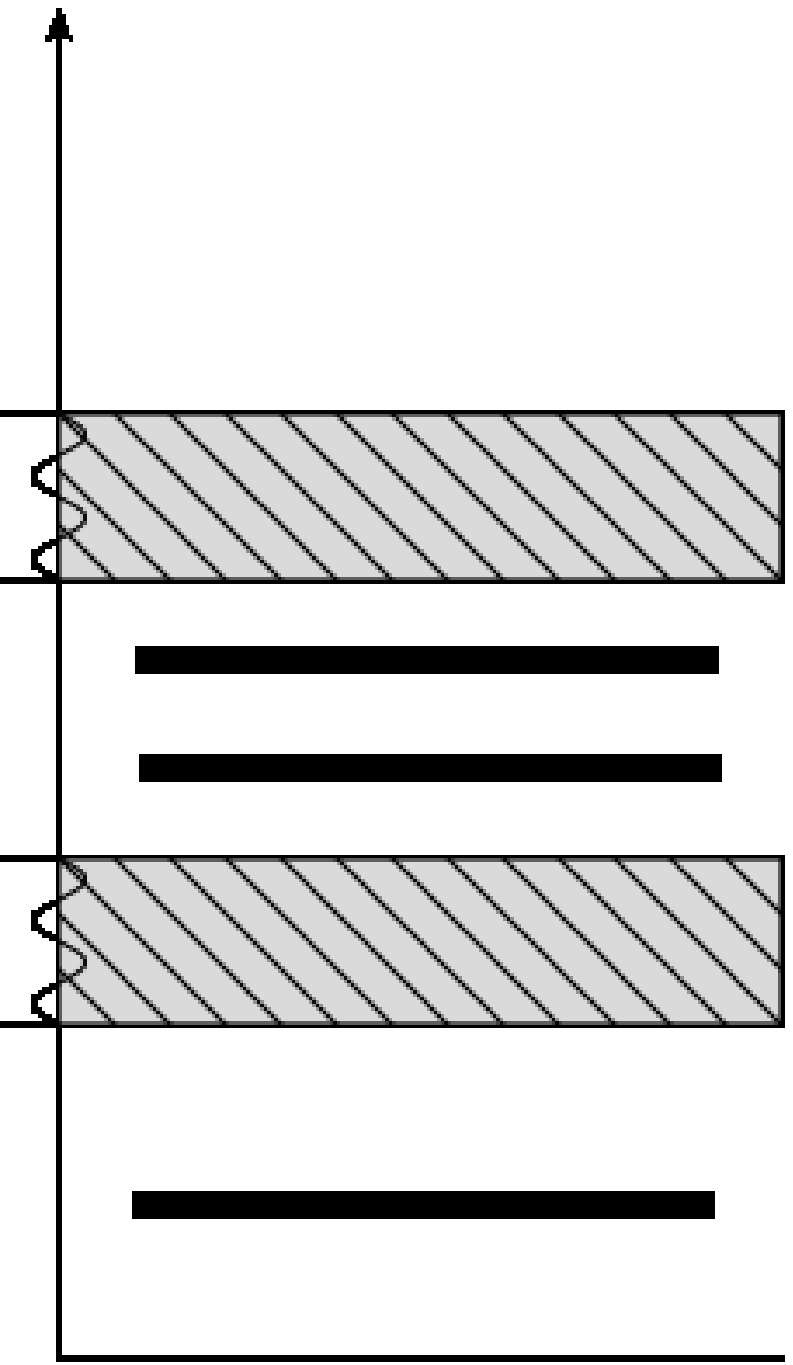,height=3cm,width=15mm}
\psfig{file=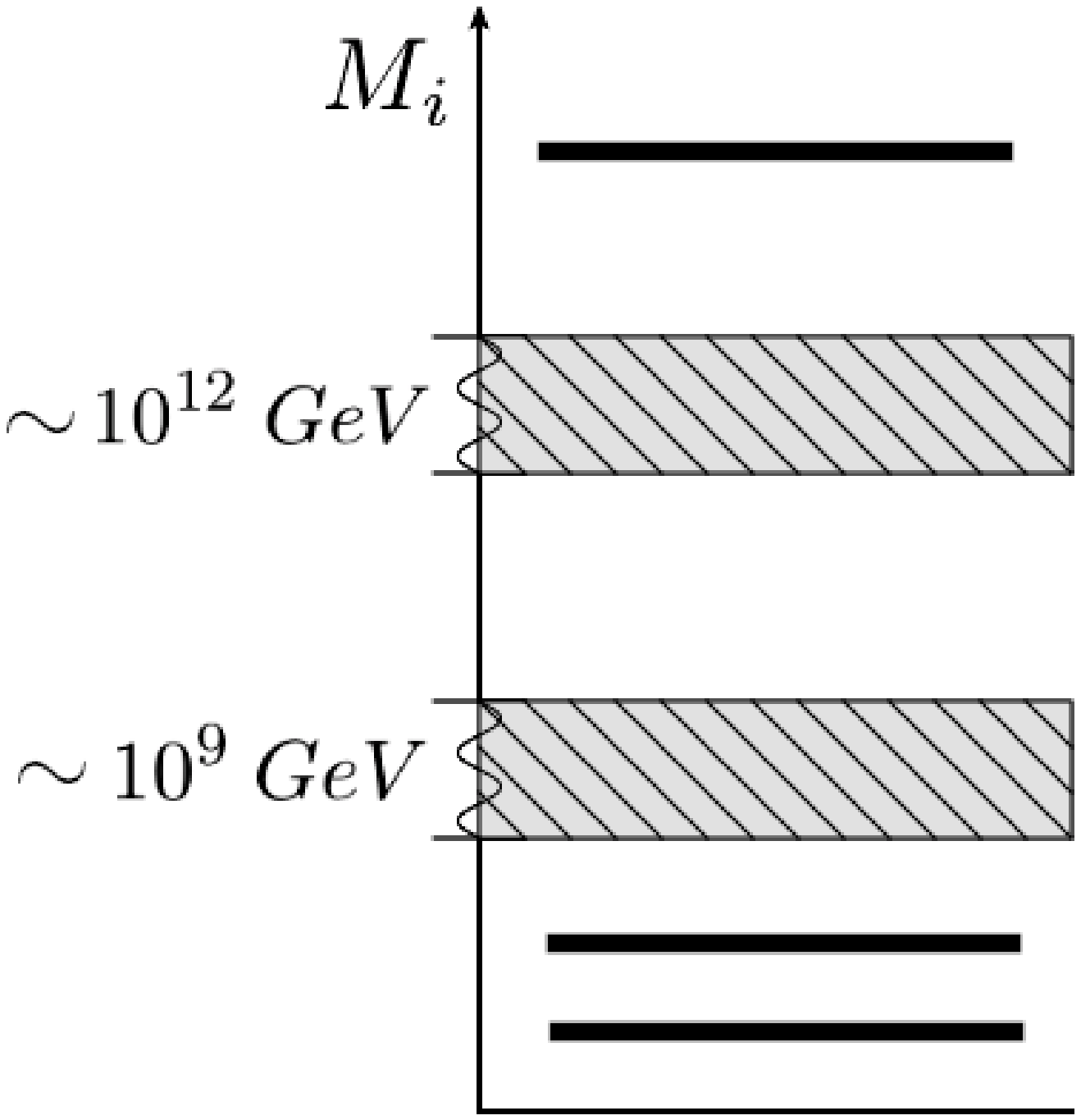,height=3cm,width=3cm}
\psfig{file=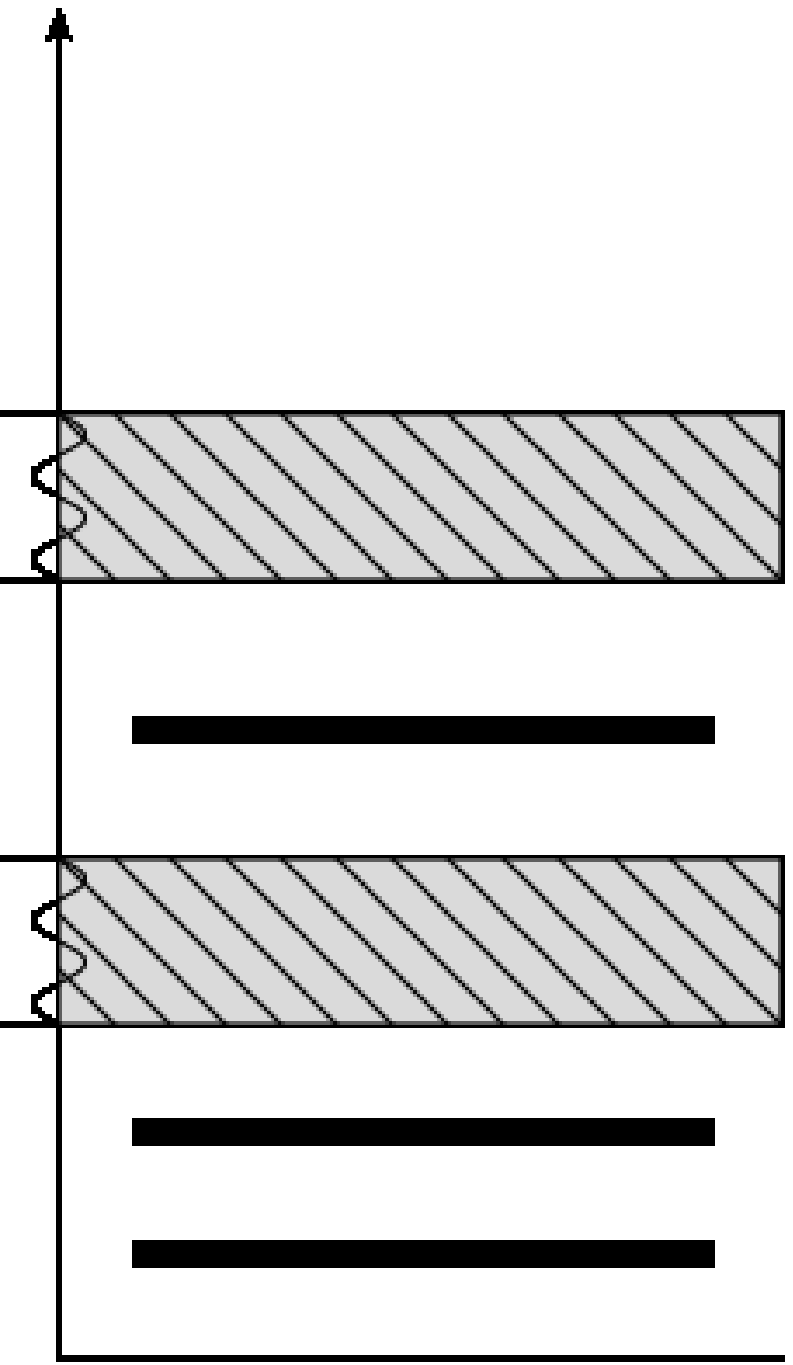,height=3cm,width=15mm}
\psfig{file=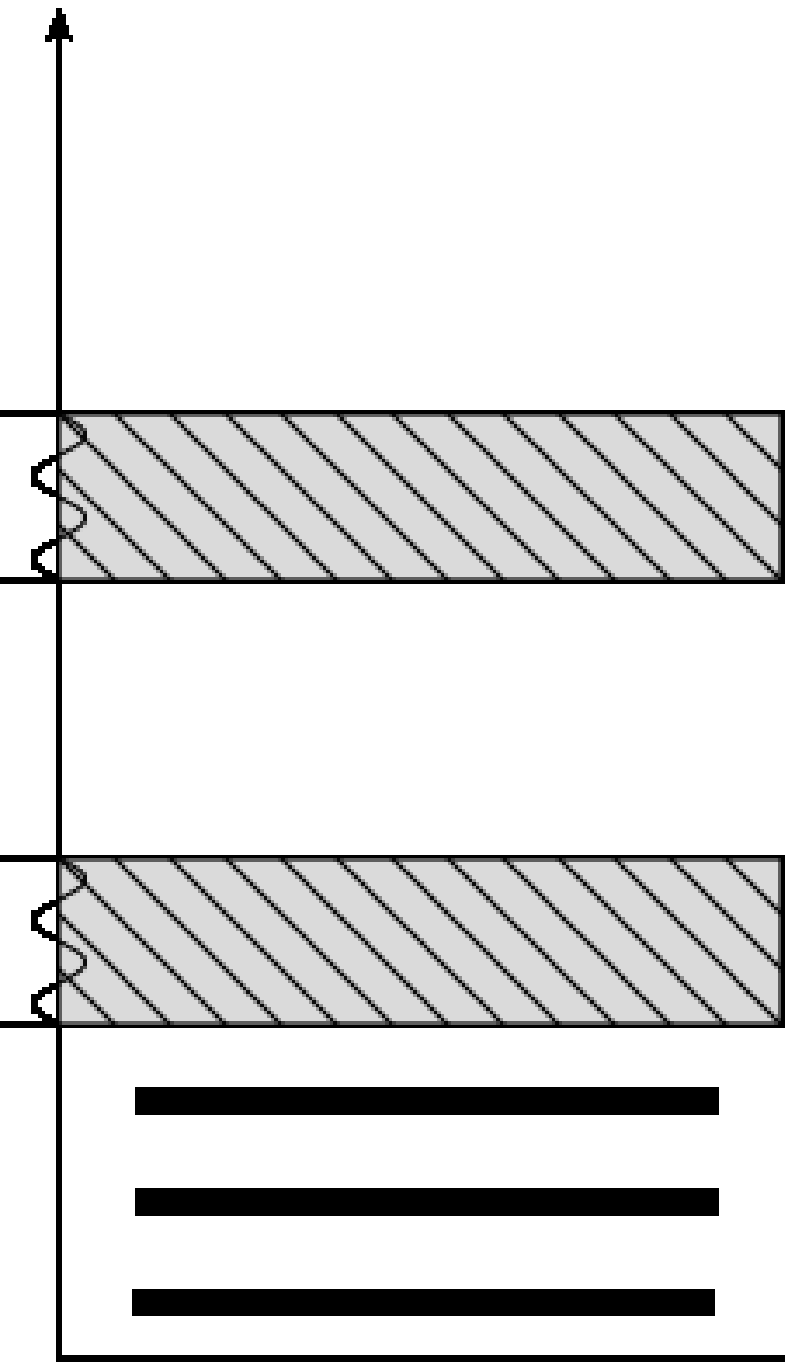,height=3cm,width=15mm}
\end{center}
\caption{The ten RH neutrino mass patterns corresponding to
leptogenesis scenarios with
different sets of classical Boltzmann equations for the
calculation of the final asymmetry \cite{problem}.}
\end{minipage}
\end{figure}

\subsubsection{The (flavoured) $N_2$-dominated scenario}

Among these 10 RH neutrino mass patterns,
for those three  where the $N_1$ washout occurs in
the three-fully-flavoured regime, for $M_1 \ll 10^{9}\,{\rm GeV}$,
the final asymmetry has necessarily to be produced by the
$N_2$ either  in the one or in the two-fully-flavoured
regime (for $M_2 \gg 10^9\,{\rm GeV}$). They have some particularly attractive
features and realise a `flavoured $N_2$-dominated scenario' \cite{vives}
(it corresponds to the fifth, sixth and seventh mass pattern in Fig.~3).

While in the unflavoured approximation the lightest RH neutrino washout
yields a global  exponential washout factor, when lepton flavour effects are
taken into account,  the asymmetry produced by the
heavier RH neutrinos, at the $N_1$ washout, gets distributed
into an incoherent mixture of charged lepton flavour eigenstates \cite{vives}.
It turns out that the $N_1$ washout in one of the three flavours is negligible, corresponding to have
at least one $K_{1\a}\lesssim 1$,
in quite a wide region of the parameter space \cite{bounds}.
In this way, accounting for flavour effects, the region of applicability  of the
$N_2$-dominated scenario  enlarges considerably, since it is not
necessary that $N_1$ fully decouples but it is sufficient that it decouples
just in a specific lepton flavour \cite{vives}. 
The unflavoured $N_2$-dominated scenario is recovered
in the limit where all three $K_{1\a}\lesssim 1$ 
and either $M_2\gtrsim 10^{12}\,{\rm GeV}$ or for 
$10^{12}\,{\rm GeV} \gtrsim M_2\gtrsim 10^{9}\,{\rm GeV}$ and $K_2 \lesssim 1$. 

Recently, it has also been realised that,
accounting for the Higgs and for the quark asymmetries, the dynamics of the flavour asymmetries
couple and the lightest RH neutrino washout in a particular flavour can be circumvented even when $N_1$ is strongly
coupled in that flavour \cite{Antusch:2010ms}.
Another interesting effect arising in the $N_2$-dominated scenario is {\em phantom leptogenesis}.
This is a pure quantum-mechanical effect that for example
allows  parts of the electron and of the muon asymmetries, the phantom terms,
to undergo a weaker washout at the production than the total asymmetry.
It has been recently shown that phantom terms
associated to a RH neutrino species $N_i$ with $M_i \gg 10^{9}\,{\rm GeV}$
are present not just in the $N_2$-dominated scenario \cite{Blanchet:2011xq}.
However, it should be noticed that phantom terms produced by the
lightest RH neutrinos
cancel with each other and thus do not contribute to the final asymmetry, though
they can induce flavoured asymmetries much larger than the total asymmetry,
something potentially relevant in active-sterile neutrino oscillations~\cite{activesterile}.

\subsubsection{Heavy neutrino flavour projection}

Even assuming a strong RH neutrino mass hierarchy, a coupled $N_1$ in all lepton flavours
($K_{1\a} \gg 1$ for any $\a$) and
$M_1\gtrsim 10^{12}\,{\rm GeV}$ (it corresponds to the first panel in Fig.~2), the asymmetry produced by the
heavier RH neutrino decays at $T\sim M_i$,  in particular by the $N_2$'s decays,
can be large enough to explain the observed asymmetry by avoiding most of the washout from 
the lightest RH neutrino
processes. This is because, in general, there is an `orthogonal' component
that escapes the $N_1$ washout \cite{bcst,Engelhard:2006yg} while the remaining
`parallel' component undergoes the  usual exponential washout.
For a mild mass hierarchy, $\d_3 \lesssim 10$,
even the asymmetry produced by the $N_3$'s decays can be large enough to
explain the observed asymmetry and escape
 the $N_1$ and $N_2$ washout.  Heavy neutrino flavour projection is also occurring
 when $10^{12}\,{\rm GeV}\gtrsim M_1 \gtrsim 10^9 \,{\rm GeV}$, in this case  
 in the $e$--$\mu$ plane.

When the effect of heavy neutrino flavour projection is taken into account jointly
 with an additional contribution to the flavoured $C\!P$ asymmetries $\ve_{2\a}$
 that is not suppressed when the heaviest RH neutrino mass $M_3\gtrsim 10^{15}\,{\rm GeV}$,
this can lead to the possibility of a dominant contribution from the next-to-lightest RH neutrinos
even in an effective  two RH neutrino model that can be regarded as  
as a limit case in a (more appealing) three RH neutrino case with 
$M_3\gtrsim 10^{15}\,{\rm GeV}$ \cite{Antusch:2010ms}.

\subsubsection{The problem of the initial conditions in flavoured leptogenesis}

As we have seen, in the vanilla scenario the unflavoured assumption reduces the problem
of the dependence on the initial conditions to simply imposing
the strong washout condition
$K_1 \gg 1$. In other words, there is a full equivalence between
strong washout and independence of the initial conditions.

When (lepton and heavy neutrino) flavour effects are considered, the situation is much more
involved and for example imposing  strong washout conditions on all flavoured
asymmetries ($K_{i\a}$) is not enough to guarantee independence of the initial conditions.
Perhaps the most striking consequence is that in a traditional
$N_1$-dominated scenario there is no condition that can guarantee
 independence of the initial conditions.  The only possibility to have independence
 of the initial conditions is represented by a tauon $N_2$-dominated scenario~\cite{problem},
 i.e. a scenario where the asymmetry is dominantly produced
 from the next-to-lightest RH neutrinos, and therefore $M_2\gg 10^{9}\,{\rm GeV}$,
 in the tauon flavour. The condition $M_1 \ll 10^{9}\,{\rm GeV}$
 is also important to have projection on the orthonormal three lepton flavour
 basis before the lightest RH neutrino washout \cite{Engelhard:2006yg}.

 \section{Density matrix formalism}

As explained in the last Section, leptogenesis is sensitive to the temperature
range at which the asymmetry is produced. This determines whether the lepton quantum states produced in RH neutrino
decays either remain coherent or undergo decoherence and get projected in flavour space before scattering in inverse
processes. Moreover, since the lepton states produced in heavy neutrino decays differ in general
from the lepton flavour eigenstates, lepton flavour
oscillations can also in principle arise in a similar way as neutrino oscillations happen in
vacuum or in a medium. In order to treat the problem of flavour oscillations and
partial loss of decoherence in a consistent way, one has to extend the classical
Boltzmann framework to account for these intrinsically quantum effects.
The formalism of the density matrix is appropriate for this purpose~\cite{Sigl:1992fn}. The
density matrix for leptons with momentum ${\mathbf p}$ is defined as
\be
\rho_{\ell}({\mathbf p})=\left(\begin{array}{cc}\langle a_{\alpha}^{\dagger}({\mathbf p})\,a_{\alpha}({\mathbf p})\rangle &
\langle a_{\beta}^{\dagger}({\mathbf p})\, a_{\alpha}({\mathbf p})\rangle \\
\langle a_{\alpha}^{\dagger}({\mathbf p})\,a_{\beta}({\mathbf p})\rangle   &
\langle a_{\beta}^{\dagger}({\mathbf p})\, a_{\beta}({\mathbf p})\rangle \end{array}\right) \, ,
\ee
with $a\,(a^{\dagger})$ denoting
the annihilation (creation) operator, i.e. it is the expectation
value (to be understood in statistical terms) of the generalised number operator.
For anti-leptons one has analogously a density matrix $\rho_{\bar{\ell}}({\mathbf p})$.
The diagonal elements of the density matrix contains
nothing else than the occupation numbers of the two flavoured leptons, and the off-diagonal
elements encode flavour correlations.

Let us consider an $N_1$-dominated scenario for simplicity. Moreover, let us consider
a momentum integrated description, introducing the matrix of lepton number densities
$N_{\ell}/R^3=g_{\ell}\int\,{d^3p\over (2\pi)^3} \, \rho_{\ell}({\mathbf p})$, where $R$ is the 
scale factor. 
Within the density matrix formalism,
the asymmetry can be calculated from the following
density matrix equation in the flavour space $\tau$--$\tau^{\bot}$
~\cite{Abada:2006fw,DeSimone:2006dd,Blanchet:2008hg,
Beneke:2010dz,Blanchet:2011xq}
\bea \label{densmateq}
{dN^{B-L}_{\a\b} \over dz} &=&
\ve^{(1)}_{\a\b}\,D_1\,(N_{N_1}-N_{N_1}^{\rm eq})-{1\over 2}\,W_1\,\left\{{\mathcal P}^{0(1)} , N^{B-L}\right\}_{\a\b}\nonumber \\ \nonumber
&+&  {\rm i}\,{{\rm Re}(\Lambda_{\t})\over H\, z}\left[\left(\begin{array}{cc}
1 & 0 \\
0 & 0
\end{array}\right),N^{\ell +\bar{\ell}}\right]_{\a\b} \\
& - & {{\rm Im}(\L_{\t})\over H\, z}\left[\left(\begin{array}{cc}
1 & 0 \\
0 & 0
\end{array}\right),\left[\left(\begin{array}{cc}
1 & 0 \\
0 & 0
\end{array}\right),N^{B-L} \right]\right]_{\a\b} \, ,
\eea
where we defined $N^{\ell +\bar{\ell}}_{\a\b} \equiv  N^{{\ell}}_{\a\b} +N^{{\bar{\ell}}}_{\a\b}$ and
where $\mathcal{P}^{0(1)}$  is a matrix projecting the
lepton quantum states along the flavour `$1$' of a lepton ${\ell}_1$
produced from the decay of a RH neutrino $N_1$ . The $C\!P$ asymmetry
matrix is a straightforward generalization of Eq.~(\ref{flavouredCP})~\cite{Abada:2006ea,Beneke:2010dz,Blanchet:2011xq}.
The real and imaginary parts of the tau-lepton self-energy are respectively given by~
\cite{Weldon:1982bn,Cline:1993bd}
\be
{\rm Re}(\L_{\t})= {f_{\t}^2\over 64} T \,  \hspace{10mm} \mbox{\rm and}
\hspace{10mm} {\rm Im}(\L_{\t})= 8\times 10^{-3}f_{\t}^2 T \, ,
\ee
where $f_{\t}$ is the tauon Yukawa coupling.
The commutator structure in the third term on the RHS of Eq.~(\ref{densmateq})
accounts for oscillations in flavour space driven by the real part of
the self energy, and the double commutator accounts for damping of the off-diagonal terms
driven by the imaginary part
of the self energy.

In order to close the system of equations, we also need an
equation for the matrix $N_{\ell +\bar{\ell}}$, which is given by
\be
{dN^{\ell +\bar{\ell}}_{\a\b} \over dz} =
-{{\rm Re}(\Lambda_{\t})\over H\, z}(\sigma_2)_{\a\b} N^{B-L}_{\a\b} -S_g \,
(N^{\ell +\bar{\ell}}_{\a\b}-2\,N_{\ell}^{\rm eq }\delta_{\a\b}) \, ,
\ee
where $S_g\equiv \Gamma_g/(Hz)$ accounts for gauge interactions. As shown in \cite{Beneke:2010dz}, this term
has the effect of damping the flavour oscillations. This can be understood by noticing that gauge
interactions force $N^{\ell +\bar{\ell}}_{\a\b}=2\,N_{\ell_1}^{\rm eq }\delta_{\a\b}$, which in turn renders the oscillatory
term Eq.~(\ref{densmateq}) negligible.

Eq.~(\ref{densmateq}) should be solved in any of the intermediate regimes where lepton states
are partially coherent. Actually the range of relevance of the density matrix equation might be more
important than previously believed. Indeed, it was found in~\cite{Beneke:2010dz} that in order to
recover the unflavoured regime, one should have masses well above $10^{13}$~GeV.

As already discussed, the contribution from heavier RH neutrinos cannot be neglected in
general. Therefore, the density matrix equation (\ref{densmateq}) should be extended to
account for such effects; this was done in~\cite{Blanchet:2011xq}, where a general equation
was presented, valid for any RH neutrino decays and any temperature range. Flavour projection effects
as well as phantom terms are readily taken into account in this framework.

\section{Limit of quasi-degenerate heavy neutrinos}

If $\delta_2 \ll 1$, the $C\!P$ asymmetries $\ve_{1,2}$
get resonantly enhanced as $\ve_{1,2}\propto 1/\d_2$ \cite{flanz,Covi:1996wh,buchplumi1}.
If, more stringently, $\d_2\lesssim 10^{-2}$, then
$\eta_B \propto 1/\delta_2$ and the degenerate limit is obtained \cite{Blanchet:2006dq}.
In this limit the lower bounds on $M_1$ and on $T_{\rm reh}$
get relaxed proportionally to $\delta_2$ and at the resonance they completely disappear
\cite{Pilaftsis:1997jf,Pilaftsis:2003gt}. The upper bound on $m_1$ also 
disappears in this extreme case. In a more realistic case where the degeneracy of the RH neutrino masses
is comparable to the degeneracy of the light neutrino masses, as typically occurring  in 
models with flavour symmetries, the upper bound $m_1 \lesssim 0.1\,{\rm eV}$ obtained in the hierarchical case
imposing the validity of Boltzmann equations,  gets relaxed to $m_1 \lesssim 0.4\,{\rm eV}$
\cite{hambyestrumia,bounds} in the case $M_3\gg M_2$, while it is basically unchanged
if $M_3=M_2$ \cite{bounds}. The difference is due to the fact that the relaxation is mainly 
to be ascribed to the extra-term in the $C\!P$ asymmetry mentioned at the end of Section 3. This 
extra-term, subdominant in the hierarchical case, can become dominant (if $M_3\gg M_2$) in the quasi-degenerate case
($\delta_2 \equiv (M_2-M_1)/M_1 \ll 1$)
and it grows with the absolute neutrino mass scale instead of being suppressed as the usual term \cite{hambyestrumia}.   
On the other hand this extra-terms vanishes exactly when $M_2=M_3$, 
a more reasonable assumption for $\d_2 \ll 1$. 

In the full three-flavour regime, the contributions from all quasi-degenerate
RH neutrinos should be taken into account and in this case the final asymmetry can be calculated as
\be
N^{\rm f}_{B-L}=\sum_{i,\alpha} \varepsilon_{i\alpha}\, \kappa\left(\sum_j K_{j\alpha}\right) \, .
\ee
Notice that, for each lepton flavour $\a$, the washout in the degenerate limit
is described by the sum of the
flavoured decay parameters for each RH neutrino species.

The simplest way to obtain a quasi-degenerate RH neutrino mass spectrum is to postulate the
existence of a slightly broken lepton number symmetry in the lepton sector~\cite{Branco:1988ex,Shaposhnikov:2006nn,
Kersten:2007vk}. Assuming the existence of
only two RH neutrinos at first for simplicity, it is possible to write down Yukawa couplings and
a Majorana mass term in a way that conserves lepton number. In the ``flavour''
basis, which we denote by a prime, one RH neutrino can be assigned lepton number +1,
and the other one -1, so that the seesaw mass matrix from Eq.~(\ref{lagrangian}) takes the form
\be\label{inverse}
M_{\nu}=\left(\begin{array}{ccc} 0 & h'_{\alpha 1}v^2 & 0\\
h^{'T}_{\alpha 1}v^2 & 0 & M \\
0 & M & 0 \end{array}\right) \, ,
\ee
which conserves lepton number. Rotating to the RH neutrino mass basis, one finds that
\be
M_{\nu}=\left(\begin{array}{ccc} 0 & h'_{\alpha 1}v^2 & {\rm i} h'_{\alpha 1} v^2\\
h^{'T}_{\alpha 1}v^2 & M & 0 \\
{\rm i} h^{'T}_{\alpha 1} v^2 & 0 & M \end{array}\right) \, .
\ee
It can be seen that the two RH neutrinos are exactly degenerate in this limit, and one
can easily show that the neutrino mass matrix $m_{\nu}$ in Eq.~(\ref{seesaw}) vanishes
identically. In other words, having non-zero neutrino masses requires a small
breaking of the lepton number symmetry, which automatically splits the two RH neutrinos
into a quasi-Dirac fermion pair.

There are different ways to implement the breaking of the
lepton number symmetry, thus generating non-zero neutrino masses. For instance,
in the above two-RH neutrino model, we can write:
\be\label{breaking}
M_{\nu}=\left(\begin{array}{ccc} 0 & h'_{\alpha 1}v^2 & \epsilon_{\alpha}v^2\\
h^{'T}_{\alpha 1}v^2 & \mu_1 & M \\
\epsilon^{T}_{\alpha}v^2 & M & \mu_2 \end{array}\right) \, ,
\ee
which implies that the full light neutrino mass matrix is given by~\cite{Gavela:2009cd,Blanchet:2009kk}
\be\label{nuinverse}
m_{\nu}\simeq v^2\left(\epsilon {1\over M} h^{'T} + h' {1\over M} \epsilon^T\right) - v^2
\left(h'{1\over M}\mu_2 {1\over M} h^{'T} \right) \, ,
\ee
proportional to the breaking parameters $\epsilon$ and $\mu_2$, as expected.
If the lepton number symmetry is broken as in the first term in Eq.~(\ref{nuinverse}), it
is referred to as `linear'~\cite{Barr:2003nn}; if the second term is at work, it is referred to
as inverse/double
seesaw mechanism~\cite{Mohapatra:1986bd}. In the following, we
will refer to these models for simplicity as
`inverse seesaw models'.
No matter how the lepton number symmetry is broken, the bottom line is that these
models fall in the category of \emph{low-scale seesaw
models}, where the size of the Yukawa couplings is not necessarily suppressed if the RH neutrino
mass scale is lowered to the electroweak scale. This can lead to interesting non-unitarity effects in neutrino
oscillation experiments~\cite{FernandezMartinez:2007ms,Antusch:2006vwa}, as well
as observable lepton flavour violating rates in experiments looking for $\mu \to e\gamma$,
$\tau \to \mu (e)\gamma$, $\mu \to eee$ or $\mu \to e$ conversion in nuclei~\cite{Pilaftsis:2005rv}.
Note that, within the orthogonal parameterisation Eq.~(\ref{casas}), the inverse seesaw model
with two RH neutrinos is obtained in the limit $|\Omega| \to \infty$~\cite{Asaka:2008bj}.

The simple model in Eq.~(\ref{inverse}) can be trivially
extended to have a third massive RH neutrino of mass still conserving lepton number.
It would have zero lepton
number and be decoupled from leptons. Without further assumptions, the other RH neutrino mass
scale is independent of $M$, and therefore it can be much lower or much higher.

Leptogenesis in the context of these low-scale seesaw models was intensely studied in recent
years. It was found in~\cite{Pilaftsis:2005rv} that it is possible to have at the same time
successful leptogenesis and low energy observable effects (beyond standard neutrino oscillation phenomenology)
such as, for example, charged lepton flavour violation processes. 
This is possible in a model with three RH neutrinos and with the help
of very large flavour effects (hence the name `resonant $\tau$-leptogenesis').

More recently, this possibility was examined in the context of a two-RH
neutrino model, and it was found that, in the limit $M_2-M_1 \gg \Gamma_{1,2}$, leptogenesis does
not allow large enough Yukawa couplings to have non-trivial consequences at low energies~\cite{Asaka:2008bj}.
This conclusion was re-examined in \cite{Blanchet:2009kk}
relaxing the requirement on the mass splitting, and allowing for more extreme quasi-degeneracies. In the
limit $M_2-M_1 \ll \Gamma_{1,2}$, it was argued that the decay parameter $K$ should be replaced by
an effective decay parameter $K^{\rm eff}_{\alpha} \propto K_{\alpha} (M_2-M_1)^2/\Gamma_1^2$
which depends explicitly on the small breaking of the lepton number
symmetry. As a matter of fact,
it is expected that the washout of lepton number vanishes in the limit
of lepton number conservation, and in \cite{Blanchet:2009kk} it was rigorously derived from the negative interference
between the two RH neutrinos exchanged in the $\Delta L=2$ process $\ell \Phi \to \bar{\ell}\Phi^{\dagger}$.

A more controversial issue is the behavior of the $C\!P$ asymmetry parameter in the limit
$M_2-M_1 \ll \Gamma_{1,2}$, which is directly related to the form of the regulator for the RH neutrino
propagator in the self-energy diagram. The reason is that the location of the pole for the RH neutrino determines
the maximum enhancement of the $C\!P$ asymmetry. Following \cite{Anisimov:2005hr}, Ref.~\cite{Blanchet:2009kk}
uses
\bea\label{CPdeg}
\varepsilon_{i\alpha} &\simeq& {1\over 8\pi (h^{\dagger}h)_{ii}}\sum_{j\neq i}
\left\{ {\rm Im}\left[
h_{\alpha i}^{\star}h_{\alpha j}(h^{\dagger}h)_{ij}\right] +
{\rm Im}\left[h_{\alpha i}^{\star}h_{\alpha j}(h^{\dagger}h)_{ji}\right]\right\} \nonumber\\
&&\times{M_j^2-M_i^2 \over (M_j^2-M_i^2)^2
+(M_i\Gamma_i -M_j\Gamma_j)^2}
\eea
with the regulator given by the difference $M_i\Gamma_i -M_j\Gamma_j$,
whereas \cite{Pilaftsis:2005rv}  finds the regulator $M_i\Gamma_i$. Within 
the inverse seesaw model considered (with two RH neutrinos)
the decay rates $\Gamma_1$ and $\Gamma_2$ are predicted to be equal
in the lepton number conserving limit. Therefore, a regulator $M_1\Gamma_1 -M_2\Gamma_2$ allows for a much
larger enhancement of the $C\!P$ asymmetry than $M_1\Gamma_1$, implying that
leptogenesis is compatible with observable lepton flavour violation rates~\cite{Blanchet:2009kk}.
The precise value of the $C\!P$ asymmetry in the regime $M_2-M_1 \ll \Gamma_{1,2}$, which is especially relevant in inverse
seesaw models, is currently still an open issue, whose resolution presumably lies beyond the classical Boltzmann approach
(see next Section).

Note that the Weinberg-Nanopoulos~\cite{Nanopoulos:1979gx} requirement, that at least
two couplings should violate lepton (or baryon)  number to have a generation of asymmetry,
is satisfied with both regulators in the inverse seesaw model.
Indeed, when four out of the five lepton-number-violating couplings in Eq.~(\ref{breaking})
are turned off, the $C\!P$ asymmetry vanishes with both regulators,
as it should~\cite{Blanchet:2009kk}. However, the limit of all couplings taken
simultaneously to zero is not well-behaved for the regulator in Eq.~(\ref{CPdeg}),
as noted in \cite{Deppisch:2010fr}.

A different scenario was considered in \cite{Antusch:2009gn} but still within the inverse seesaw framework. There,
a third almost decoupled RH neutrino was added, with a mass $M_1\ll M$. 
Lepton number violation was included in the coupling of the lightest
RH neutrino, such that it can decay producing a lepton asymmetry. Therefore, in this case, the quasi-degenerate
pair of RH neutrinos is not responsible for the generation of asymmetry. Nonetheless, their large couplings to leptons
(leading for instance to non-unitarity effects in neutrino oscillations) imply that the flavoured $C\!P$ asymmetry, more precisely
the lepton number conserving part in Eq.~(\ref{flavouredCP}), can be large even for TeV-scale RH neutrino masses.
However, this scenario of `non-unitarity driven leptogenesis' has intrinsically large lepton
flavour violating interactions that lead to flavour equilibration~\cite{AristizabalSierra:2009mq}. Using the
cross-sections for flavour violating interactions obtained in~\cite{Pilaftsis:2005rv}, one finds that
the asymmetry cannot be generated in the right amount if $M_1\lesssim 10^8$~GeV~\cite{Antusch:2009gn}. 
However, these cross-sections were found recently to be significantly more suppressed than in~\cite{Pilaftsis:2005rv}, 
leading to an interesting
relaxation of the bounds down to $M_1\gtrsim 10^6$~GeV~\cite{Racker:2012vw}.

It is worth mentioning that leptogenesis was investigated in the context of the inverse seesaw model also when
lepton number is exactly conserved \cite{GonzalezGarcia:2009qd}.
In this case, the observed baryon asymmetry can be generated
if leptogenesis occurs during the electroweak phase transition, when the sphaleron rate progressively
goes out of equilibrium. The lepton flavour asymmetries, generated exclusively thanks to flavour effects (again the second
term in Eq.~(\ref{flavouredCP})),
are then converted into a baryon asymmetry before total washout.

Another example where leptogenesis occurs in the resonant regime
is radiative leptogenesis~\cite{GonzalezFelipe:2003fi,Branco:2005ye}. There, RH neutrinos are assumed exactly
degenerate at some high scale (for instance the GUT scale), and
small RH neutrino mass splittings are generated by the running of the seesaw parameters.

In these low-scale seesaw models with TeV-scale RH neutrinos,
it is natural to wonder whether some interesting signatures could be observed at collider experiments such as the LHC.
Unfortunately, it seems that within the simplest type-I seesaw, the prospects are
rather dim~\cite{Kersten:2007vk,Ibarra:2011xn}.
The main problem is that the mixing of RH neutrinos with light neutrinos has strong upper
limits from rare (lepton-flavour-violating) decays~\cite{FernandezMartinez:2007ms}, which prevents an
important production of RH neutrinos at the LHC.
However, in extended models such as Type II seesaw, Type III  seesaw, left-right symmetric
models, or simply an extra $U(1)_{B-L}$, 
the prospects are much more encouraging (see \cite{hambye} in this Issue).

\section{Improved kinetic description}

We have already discussed the  density matrix formalism, which goes beyond the traditional
kinetic treatment with Boltzmann (rate) equations. This Section is devoted to other
kinetic effects that are important both for an estimation of the
theoretical uncertainties in the calculation and for a better conceptual understanding of
the minimal leptogenesis framework.

\subsection{Momentum dependence}

Within the vanilla scenario, the final asymmetry is computed by solving classical Boltzmann equations
for the RH neutrino and lepton \emph{number densities}, so-called rate equations. These
are obtained from the Boltzmann equations for  the \emph{distribution
function} integrating over momenta with some approximations (see below).
One can then wonder what is the theoretical error introduced by this integrated description.

Given a particle species $X$, the number density is obtained by integrating
the distribution function over momentum,
\be
n_X ={g_X\over(2\pi)^3}\, \int \, d^3 \, p_X \, f_X \, ,
\ee
where $g_X$ is the number of degrees of freedom of particle $X$.
For leptogenesis with decays and inverse decays,
the system of Boltzmann equations (one for the RH neutrino and one for lepton number)
in the expanding Friedmann-Robertson-Walker Universe
is given by
\bea
{\partial f_N\over \partial t} - |{\bf p}_N| H {\partial f_N\over \partial |{\bf p}_N|}&=& \mathcal{C}_D[f_N]\\
{\partial f_{\ell-\bar{\ell}}\over \partial t} - |{\bf p}_{\ell}| H
{\partial f_{\ell-\bar{\ell}}\over \partial |{\bf p}_{\ell}|}&=& \mathcal{C}_D[f_{\ell-\bar{\ell}}] \, ,
\label{momentumke}
\eea
where the collision integrals on the right-hand side are defined as
\bea \fl
\mathcal{C}[f_A, A \leftrightarrow B\, C] &=& {1\over 2 E_A}\int {d^3 p_B \over 2E_B (2\pi)^3}
{d^3 p_C \over 2E_C (2\pi)^3} (2\pi)^4 \delta^4(p_A - p_B - p_C) \nonumber\\
&&\times \left[ f_B f_C(1-f_A)|\mathcal{M}( B\,C\to A)|^2 - \right.\\
&&\left. f_A(1-f_B)(1-f_C)|\mathcal{M}(A\to B\,C)|^2\right]\nonumber \,.
\eea
Note that the double-counting problem is
solved here in the same way as for the integrated Boltzmann equations, by consistently including
the resonant part of the $\Delta L=2$ scatterings. 
In order to recover the usual Boltzmann equations~(\ref{dlg1})--(\ref{unflke}),
one has to introduce three approximations:
(i) kinetic equilibrium for the RH neutrinos, which can be expressed as $f_N/f_N^{\rm eq} = n_N/n_N^{\rm eq}$;
(ii) Maxwell-Boltmann distributions for RH neutrinos, leptons and Higgs fields;
(iii) neglect Pauli blocking and Bose enhancement factors.

RH neutrinos are only coupled to the thermal bath via their Yukawa couplings. It is therefore
clear that in the weak washout regime, $K_1\ll 1$, the assumption of kinetic equilibrium is not a very good one,
and, indeed, it was found that the lepton asymmetry computed with the above equations can differ by
up to 50\% in the weak
washout regime compared to the usual treatment with integrated equations~\cite{Basboll:2006yx,HahnWoernle:2009qn,
Garayoa:2009my}.
However, as expected, in the strong washout regime, the above-mentioned approximations
are very good and the integrated rate equations can be used safely.

\subsection{Non-equilibrium formalism}

Leptogenesis is an intrinsic non-equilibrium problem. One of the Sakharov's conditions is indeed that
a departure from thermal equilibrium is necessary to produce the baryon asymmetry. Therefore,
it does not come as a surprise that recently a huge effort~\cite{Buchmuller:2000nd,desimone,qke,qke2,garny,beneke,
Beneke:2010dz,Garny:2011hg,Garbrecht:2011aw}
was made to understand leptogenesis
within non-equilibrium quantum field theory, also known as the closed-time-path (CTP) or
Keldysh-Schwinger formalism. This more rigorous, though far more complex, approach has the advantage
of taking into account quantum effects that are completely missed by the usual approach,
like memory effects and off-shell effects. Moreover, it allows the straightforward inclusion
of flavour oscillations and decoherence~\cite{Beneke:2010dz}, it has the advantage of being able to consistently
account for finite density corrections, and it has no double counting problem.

Concretely, in the non-equilibrium framework,
one needs to find the equations of motion for the two-point correlation functions
(i.e. the propagators or Green's functions)  of RH neutrinos
and leptons from the general Schwinger-Dyson equation on the CTP,
\bea
S_{\ell}^{-1}(x,y)&=& S^{-1}_{\ell 0}(x,y) -\Sigma_{\ell}(x,y)\, ,\\
S^{-1}(x,y)&=& S^{-1}_{0}(x,y) -\Sigma_{N}(x,y)\, ,
\eea
which are obtained from the variational principle on the effective action, a functional of the
full propagators $\Delta_{\phi}$, $S_{\ell}$ and $S$, for the Higgs, lepton and RH neutrino, respectively.
In the above equations, the subscript $0$ denotes the free propagators, and $\Sigma_{\ell}$ and
$\Sigma_N$ are the self-energies for the leptons and RH neutrinos, respectively.
Note that the self-energies are themselves functions of the propagators. For instance, the one-loop
lepton self-energy depends on the RH neutrino and Higgs propagators:
\be
\Sigma_{\ell}^{\alpha \beta}(x,y)=-h_{\alpha i}h_{j \beta}^{\dagger} P_R S^{ij}(x,y)P_L \Delta_{\phi}(y,x) \, .
\ee
It usually proves convenient
to decompose any two-point function $D(x,y)$ into a \emph{spectral}, $D_{\rho}$, and a \emph{statistical}
component, $D_F$:
\be
D(x,y)=D_F(x,y)-{i\over 2} {\rm sgn}(x^0-y^0)D_{\rho}(x,y) \, .
\ee
Convoluting the Schwinger-Dyson equations
with the full propagator, we finally arrive at a system of two coupled integro-differential
equations, the so-called Kadanoff-Baym equations:
\bea
i\slashed{\partial}_x {S_{\ell}}_F^{\alpha\beta}(x,y)&=&\int_0^{x^0} d^4 z {\Sigma_{\ell}}_{\rho}^{\alpha\gamma}(x,z)
{S_{\ell}}_F^{\gamma\beta}(z,y) \nonumber\\
&& -\int_0^{y^0} d^4 z {\Sigma_{\ell}}_{F}^{\alpha\gamma}(x,z)
{S_{\ell}}_{\rho}^{\gamma\beta}(z,y)\, , \\
i\slashed{\partial}_x {S_{\ell}}_{\rho}^{\alpha\beta}(x,y)&=&\int_{y^0}^{x^0} d^4 z {\Sigma_{\ell}}_{\rho}^{\alpha\gamma}(x,z)
{S_{\ell}}_{\rho}^{\gamma\beta}(z,y) \, .
\eea
The corresponding equations for the RH neutrino two-point function are obtained by changing $\slashed{\partial}_x \to
\slashed{\partial}_x -M$, and $S^{\alpha \beta} \to
S^{ij}$, where $\alpha, \beta= e,\mu,\tau$ are lepton flavours, and $i,j=1,2,3$ are RH neutrino flavours.
It can be noticed that the Kadanoff-Baym equations contain an integration over the entire history of the system, a `memory'
integral, which encodes all previous interactions with momentum and spin correlations. An attempt
of studying memory effects in the context of leptogenesis~\cite{desimone}
has found that large effects could arise in the resonant limit in the weak washout regime.
In order to recover a Markovian description of
the system, characterized by uncorrelated initial states at every timestep, one has to perform a gradient expansion
(for an alternative approach, see~\cite{qke2}),
relying on the fact that the microscopic timescale $t_{\rm mic}\sim 1/M_i$ is much smaller than the macroscopic timescales
$t_{\rm mac}\sim 1/\Gamma_i,1/H$. This is also known as the molecular chaos approximation.

In leptogenesis, one needs to compute the evolution of the lepton number density. The latter
is given by the average expectation value of the zeroth component of the lepton number current,
given by
\be
j^{\mu}_{L\alpha\beta}(x)=-{\rm tr}\left[\gamma^{\mu}S_{\ell \alpha\beta}(x,x)\right] \, .
\ee
One then obtains for the lepton number density
\bea
n_{L\alpha\beta}(t)&=& i\int {d^3 p\over (2\pi)^3} \int_0^t dt' \int_0^{t'} dt'' {\rm tr}
\left[{{\Sigma_{\ell}}_{\rho}}_p^{\alpha\beta} (t',t'') {{S_{\ell}}_F}_p(t'',t') \right.\nonumber \\
&& \left. -{{\Sigma_{\ell}}_F}_p^{\alpha \beta} (t',t'') {{S_{\ell}}_{\rho}}_p(t'',t')\right] \, ,
\eea
after switching to momentum space. From this master equation, one then needs to
input the equilibrium expression for the lepton and
Higgs propagators, as well as the non-equilibrium Majorana neutrino propagator.
In order to obtain a Boltzmann-like equation, two further simplifications are required.
First, the quasi-particle ansatz, also known as the on-shell approximation, which states:
\be
D_{\rho}(X,p)=2\pi\, {\rm sgn}(p^0)\delta(p^2-m^2) \, ,
\ee
where $X\equiv (x+y)/2$ is the central coordinate. Then, one can express
the statistical propagator for the occupation number using the so-called Kadanoff-Baym ansatz,
\be
D_F(X,p)=\left[f(X,p)+{1\over 2}\right]D_{\rho}(X,p) \, ,
\ee
which is chosen such that, in equilibrium, the correct \emph{fluctuation dissipation relation} between $D_F^{\rm eq}(p)$ and $D_{\rho}^{\rm eq}(p)$ 
is automatically obtained (see \cite{qke} for more details).
With these approximations, one arrives at a Boltzmann equation which includes finite density
effects. This equation would be similar to the kinetic equation~(\ref{momentumke}), with however
the important new property:
\be\label{property}
|\mathcal{M}( N\to \ell \Phi)|^2 = |\mathcal{M}( \ell \Phi \to N)|^2 =
|\mathcal{M}_0( N\leftrightarrow \ell \Phi)|^2 (1+\varepsilon(p,T)) \,,
\ee
where subscript `0' denotes tree level, and~\cite{garny}
\be
\varepsilon (p, T)=\varepsilon \times \left(1+\int {d \Omega\over 4\pi}[f_{\phi}^{\rm eq}(E_1)
-f_{\ell}^{\rm eq}(E_2)]\right) \,
\ee
where $\varepsilon$ is the $C\!P$ asymmetry defined in Eq.~(\ref{CPas}), and
$E_{1,2}={1\over 2} \left[(M^2+ p^2)^{1\over 2} \pm p \cos\theta\right]$.
Note that the above property in Eq.~(\ref{property}) explicitly avoids
the double counting problem, which plagues the momentum-dependent description of the
last Subsection (as well as the usual integrated one).
As it can be seen, the
$C\!P$ asymmetry parameter now includes finite density effects via a dependence on the Higgs and lepton
distribution functions. This result
agrees with that one found using thermal field theory in the real time formalism, when the right convention
is used \cite{garny}, as well as in the imaginary time formalism as recently obtained~\cite{Kiessig:2011fw}.
It, however, disagrees with an earlier work based on thermal field theory in the real time formalism, where
a term quadratic in the distribution functions was found~\cite{giudice}.

One should then wonder whether the non-equilibrium formalism described above leads to substantial
differences in the leptogenesis predictions for the baryon asymmetry. It turns out
that the modifications are very
important (up to an order of magnitude) in the regime where $M_i/T\ll 1$, but they tend to vanish
in the non-relativistic regime $M_i/T \gg 1$. In other words, major changes are expected mainly in the
weak washout regime for a vanishing initial RH neutrino abundance, because in this case the asymmetry is produced
in both regimes: the initial asymmetry is produced with the wrong sign at early times and is compensated
by the right-sign asymmetry at later times. This was confirmed in~\cite{beneke}, where it was found
that the sign of the final asymmetry could even get changed by finite density effects.
In all other cases, corrections are quantitatively small, in particular in the strong washout regime,
$K_i\gg 1$,
where the asymmetry is produced exclusively in the non-relativistic regime~\cite{garny,beneke}.


\subsubsection{Resonant limit}

It is interesting to study whether the above quantum kinetic formalism can provide some insight
onto the $C\!P$ asymmetry produced
in the extreme quasi-degenerate limit, $|M_1 -M_2|\ll M_{1,2}$. This formalism includes off-shell
effects, as well as the proper inclusion of coherent
transitions $N_i \to N_j$, both of which should be important when $|M_1 -M_2|\sim \Gamma_{1,2}$. This
problem attracted some attention recently~\cite{Garny:2011hg,Garbrecht:2011aw}.

In \cite{Garny:2011hg} it was found that the enhancement of the $C\!P$ asymmetry
derived within the Kadanoff-Baym (KB) formalism, 
\be
R^{\rm KB}={M_1 M_2 (M_2^2-M_1^2)\over (M_2^2-M_1^2)^2+(M_1\Gamma_1+M_2\Gamma_2)^2} \,  ,
\ee
differs from the Boltzmann result, 
\be
R^{\rm BE}={M_1 M_2 (M_2^2-M_1^2)\over (M_2^2-M_1^2)^2+(M_1\Gamma_1-M_2\Gamma_2)^2} \, ,
\ee
specifically because of a contribution from coherent RH neutrino oscillations. 
This would prevent any additional enhancement when $\Gamma_1 \sim \Gamma_2$, as predicted by
the inverse seesaw model. In such a case, the new region in the parameter space of successful
leptogenesis explored in~\cite{Blanchet:2009kk} is not available.

\subsubsection{Adding flavour}

We saw in Section 5 how the first kind of `quantum' effects were included,
namely flavour oscillations and decoherence, by switching from classical Boltzmann equations
for lepton number densities to evolution equations for the density matrix. For the lightest
RH neutrino and two relevant lepton flavours, we presented Eq.~(\ref{densmateq}). The CTP
formalism can also account for a flavour matrix structure, through the
lepton propagators and self-energies. Reassuringly, the structure of
Eq.~(\ref{densmateq}) was also found within this formalism~\cite{Beneke:2010dz}.

\section{Other corrections}

\subsection{Thermal effects}

Leptogenesis occurs in the very hot thermal bath of the early Universe. Leptons and Higgs fields
have very fast interactions with the thermal bath due to their gauge couplings. This gives them an
effective mass which is proportional to temperature~\cite{Weldon:1982bn}. This effective mass allows processes
which were otherwise kinematically forbidden to occur, such as $C\!P$-violating Higgs decays to
RH neutrinos, when $T\gg M$. Thermal effects have therefore a direct impact on leptogenesis, and the
first study to try and quantify them \cite{giudice} employed a real time formalism and
hard thermal loop resummation. It was
found that the $C\!P$ violating parameter has a strong temperature dependence, and
that in the weak washout regime, there
could be important differences with the zero-temperature treatment. On the other hand, in the strong
washout regime, $K\gg 1$, the usual results from a vacuum calculation are recovered with
a good accuracy. Recently, a new study came out using the imaginary time formalism~\cite{Kiessig:2011fw},
and the $C\!P$
asymmetry parameter was found to differ from~\cite{giudice}, and to agree with more recent attempts
with non-equilibrium quantum field theory (see previous Section).

\subsection{Spectator processes}

Chemical equilibrium holds among Standard Model particles in the early Universe above the electroweak phase
transition thanks to gauge interactions. On the other hand, Yukawa interactions only force some new conditions
(between left- and right-handed fermions)
when the temperature is high enough, depending on the size of the Yukawa coupling. If one imposes
additionally hypercharge
neutrality and the effect of electroweak and strong sphaleron equilibrium, one arrives at a set
of relations among the chemical potentials (or among the asymmetries) of leptons, Higgs and
baryons~\cite{Nardi:2005hs}.
These processes, although not directly involved in the leptogenesis process, hence the name `spectator
processes'~\cite{Buchmuller:2001sr},
have an indirect effect
on the final asymmetry through a modified washout. The first effect is the inclusion of the Higgs asymmetry
as a new contribution to the washout. The second effect is that the asymmetry originally produced in leptons of a
particular flavour gets redistributed into the other flavours following precise relations~\cite{bcst}.
This means
that the set of Boltzmann equations to solve for the different lepton flavours are now coupled to each other
via a flavour coupling matrix.

Within the $N_1$-dominated scenario,
the overall effect of spectator processes is usually subdominant compared to flavour effects, but it can change
the final result by as much as 40\% depending on the temperature at which the asymmetry
is produced~\cite{Nardi:2005hs,JosseMichaux:2007zj}.
Supersymmetry has new degrees of freedom and new constraints can be derived. A full study was performed recently
in~\cite{Fong:2010qh} and the overall effect was found to be again of order one.
Within the $N_2$-dominated scenario, however, potentially much bigger effects are possible \cite{Antusch:2010ms}.

\subsection{Scattering processes}

It can be shown that leptogenesis is well-described in the strong washout regime, $K\gg 1$, by just
decays and inverse decays. The reason is that in this regime
the asymmetry is produced at relatively late times, at $T\ll M$, with no dependence on the dynamics
happening at $T\gtrsim M$ when scattering
processes are important. However, they should be included
in the weak washout regime for initial vanishing RH neutrino abundance. For instance, $\Delta L=1$ Higgs-mediated
scatterings involving top quarks, such as $\ell N \leftrightarrow Q_3 \bar{t}$, contribute to the washout
and the $C\!P$ asymmetry~\cite{Abada:2006ea}, and their inclusion is crucial to have a correct estimation
of the final asymmetry in the weak washout regime~\cite{Nardi:2007jp}. Scatterings involving gauge
bosons, such as $\ell N\leftrightarrow \Phi^{\dagger}A$, have also been included, especially their contribution
to the $C\!P$ asymmetric source term~\cite{Fong:2010bh}. It was found that the factorization of the $C\!P$ asymmetry
from decays and scatterings involving top quarks does not happen with scatterings involving gauge bosons.
In particular, there is a new source of lepton-number-conserving $C\!P$ asymmetry.

\section{Testing new physics with leptogenesis}

The seesaw mechanism with three RH neutrinos extends the Standard Model by introducing eighteen new parameters.
On the other hand, low-energy
neutrino experiments can only potentially test the nine parameters in the low-energy
neutrino mass matrix $m_{\nu}$. Nine high-energy parameters, those characterising the properties
of the three RH neutrinos, the three masses and the six parameters
encoded in the seesaw orthogonal matrix of Eq.~(\ref{casas}),
basically fixing, together with the light neutrino masses, the three lifetimes and the three total $C\!P$ asymmetries,
are not tested by low-energy neutrino experiments.
Quite interestingly, the requirement of successful leptogenesis,
\be
\eta_B(m_{\nu},\Omega,M_i)=\eta_{B}^{\rm CMB}   \,  ,
\ee
provides an additional constraint on a combination
of both low-energy neutrino parameters and high-energy neutrino parameters.
However, just one additional constraint would not seem sufficient to over-constrain the parameter
space leading to testable predictions.
In spite of this observation, as we have seen, in the vanilla leptogenesis scenario
one can derive an upper bound on the neutrino masses. The reason is that, within this scenario,
the dependence of $\eta_B$ on the six parameters related to the properties of the two heavier RH neutrinos
cancels out.  In this way the asymmetry depends on a reduced
subset of high-energy parameters (just three instead of nine).
At the same time, the final asymmetry gets strongly suppressed when
the absolute neutrino mass scale it is larger than the atmospheric neutrino mass scale.
For all these reasons, by maximising the final asymmetry over the high energy parameters and
by imposing successful leptogenesis,  an upper bound on the neutrino masses is found.

When flavour effects are considered, the vanilla leptogenesis
scenario holds only under very special conditions, as we have seen.
In general, the final asymmetry depends also on the parameters in the leptonic mixing matrix.
Therefore, accounting for flavour effects,
one could naively hope to derive definite predictions on the leptonic mixing matrix as well,
in addition to the upper bound on the absolute neutrino mass scale.
However,  the situation is quite different when flavour effects are taken into account.
This is because the final asymmetry depends, in general, also on the six parameters
related describing the two heavier RH neutrino properties and
that were cancelling out in the calculation of the final asymmetry in the vanilla
scenario, and this goes at the expense of predictability.

For this reason, in a general scenario with three RH neutrinos and flavour effects included, it is not possible
to derive any prediction on low-energy neutrino parameters.
As we discussed, even whether the upper bound $m_i\lesssim 0.1$~eV on neutrino masses still holds
or not is an open issue and a precise value seems to depend on a precise account of 
many different subtle effects and in particular it necessarily requires a density matrix formalism. 

In order to gain predictive power, two possibilities have been explored in the past years.

A first possibility is to consider non-minimal scenarios giving rise to additional phenomenological constraints.
For example, as discussed in Section 6, the inverse seesaw model, which technically can be still regarded as part
of the minimal type I seesaw model, allows for non-trivial phenomenologies at low energy beyond standard neutrino
oscillations, such as non-unitarity effects and observable lepton flavour violation. An experimental
observation of any of these signatures would be of great value to reduce the freedom in
the choice of seesaw parameters.
In recent years, during the Large Hadron Collider era,  it has been also intensively
explored the possibility that, within a non-minimal version of the seesaw mechanism,
one can have successful low scale leptogenesis together with collider signatures.
It has also been noticed that
in the supersymmetric version of the seesaw, the branching ratios of lepton-flavour-violating processes
or electric dipole moments are typically enhanced, and hence the existing experimental bounds
further constrain the seesaw parameter space \cite{Pascoli:2003uh,Dutta:2003my}.

A second possibility is to search for a reasonable
scenario where the final asymmetry depends only on a reduced set of independent parameters
over-constrained by the successful leptogenesis condition, as with the vanilla scenario. 
From this point of view, the account of
flavour effects has opened  very interesting new opportunities or even re-opened old attempts
that fail within a strict unflavoured scenario. Let us briefly discuss some
of the main ideas that have been proposed within this second possibility, the first one being
covered elsewhere in this Issue~\cite{hambye}.

\subsection{Two-RH neutrino model}

A  phenomenological possibility that attracted great attention is the two-RH
neutrino model \cite{Frampton:2002qc},
where the third RH neutrino is either absent or
effectively decoupled in the seesaw formula.
This necessarily happens when $M_3\gg 10^{14}\,{\rm GeV}$, implying that the
lightest left-handed neutrino mass $m_1$ has to vanish. It can be shown that the number of parameters
decreases from 18 to 11 in this case.
In particular the orthogonal matrix is parameterised in terms of just one complex angle.

In leptogenesis the two-RH neutrino model has been traditionally considered as a sort of benchmark case
for the $N_1$-dominated scenario, where the final asymmetry is dominated by the contribution
from the lightest RH neutrinos \cite{Abada:2006ea,Blanchet:2006dq,petcovmolinaro}. However, recently, as we anticipated already, it has been
shown that there are some regions in the one complex angle parameter space
that are $N_2$-dominated~\cite{Antusch:2011nz} and that
correspond to so-called light sequential dominated models \cite{King:2003jb}.

It should be said that even though the number of parameters is highly reduced,
in a general two-RH neutrino model it is still not possible to make predictions on the low-energy neutrino
parameters. To this extent, one should further reduce the parameter space, for example assuming texture zeros
in the neutrino Dirac mass matrix.

\subsection{$SO(10)$-inspired models}

 In order to gain  predictive power, one can
impose conditions within some model of new physics embedding the seesaw mechanism.
An interesting example is represented by the `$SO(10)$-inspired leptogenesis scenario' 
\cite{buchplumi,Branco:2002kt,Akhmedov:2003dg},
where $SO(10)$-inspired conditions are imposed on the neutrino Dirac mass matrix $m_D$.
In the basis where the charged leptons mass matrix and the Majorana mass matrix are diagonal,
in the bi-unitary parametrisation, one has $m_D = V_L^{\dagger}\,D_{m_D}\,U_R$,
where $D_{m_D}\equiv {\rm diag}({\l_1,\l_2,\l_3})$ is the diagonalised neutrino Dirac mass matrix
and the mixing angles in $V_L$ are of the order
of the mixing angles in the CKM matrix $V_{CKM}$.
The $U_R$ and three $M_i$ can then be calculated from $V_L$, $U$ and $m_i$,
since the seesaw formula Eq.~(\ref{seesaw}) directly leads to the Takagi factorisation of
$M^{-1} \equiv D^{-1}_{m_D}\,V_L\,U\,D_m\,U^T\,V_L^T\,D^{-1}_{m_D}$,
or explicitly $M^{-1} = U_R\,D_M^{-1}\,U_R^T$.

In this way the RH neutrino masses and the matrix $U_R$ are expressed in terms of the
low-energy neutrino parameters, of the eigenvalues $\l_i$ and of the parameters in $V_L$.
Typically one obtains a very hierarchical spectrum $M_1 \sim 10^{5}\,{\rm GeV}$ and $M_{2}\sim 10^{11}\,{\rm GeV}$,
the asymmetry produced from the lightest RH neutrino decays is by far unable to explain the
observed asymmetry \cite{Branco:2002kt}. 

However,  when the $N_2$ produced asymmetry is taken into account,
successful ($N_2$-dominated) leptogenesis  can be attained \cite{SO10lep1}. 
In this case, imposing the leptogenesis bound
and considering that the final asymmetry does not depend on $\l_1$ and on $\l_3$, one obtains
constraints on all low-energy neutrino parameters, which have some dependence on the
parameter $\l_2$ typically parameterised in terms of $\alpha_2\equiv \l_2/m_c$, where $m_c$
is the charm quark mass. Some examples of the constraints on the low-energy neutrino parameters
are shown in Fig.~3. They have been obtained scanning over the $2\sigma$ ranges of the allowed values of the
low-energy parameters and over the parameters
in $V_L$ assumed to be $I< V_L < V_{CKM}$ and
for three values of $\alpha_2=5,4, 1$.
It is particularly interesting that when the independence of the initial conditions
is imposed, negative values of $J_{CP}$ seem to be favoured \cite{prep} establishing
a connection between the sign of $J_{CP}$ and of the matter-antimatter asymmetry.

A supersymmetric version of this scenario including the renormalization group evolution
of all the relevant couplings was also studied in~\cite{Blanchet:2010td}, and
including a type II contribution to the seesaw mechanism
from a triplet Higgs in left-right symmetric models in~\cite{Abada:2008gs}.
\begin{figure}
\begin{center}
         \mbox{\epsfig{figure=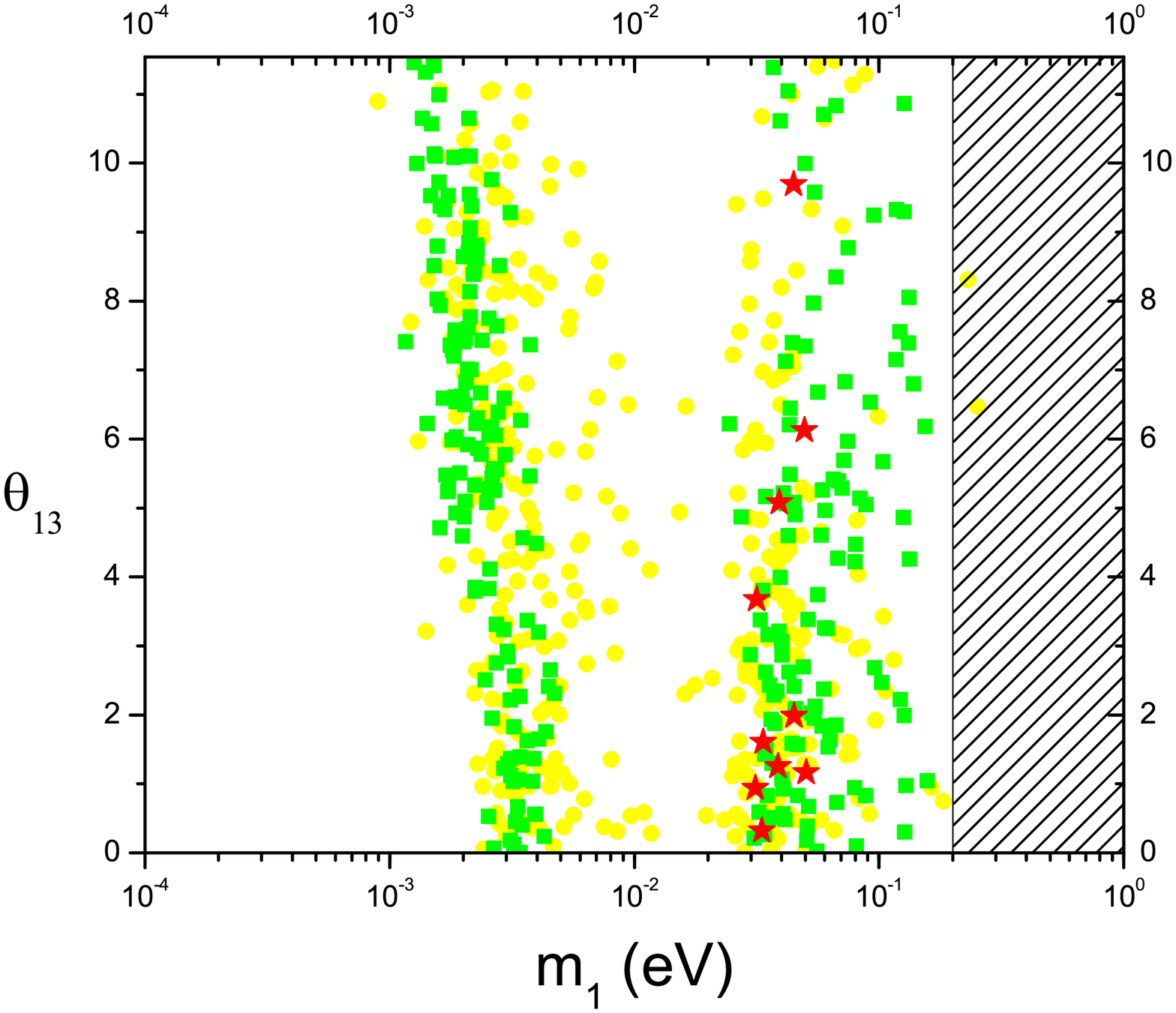,width=55mm,height=55mm}}
         \vspace*{-20pt}
        \mbox{\epsfig{figure=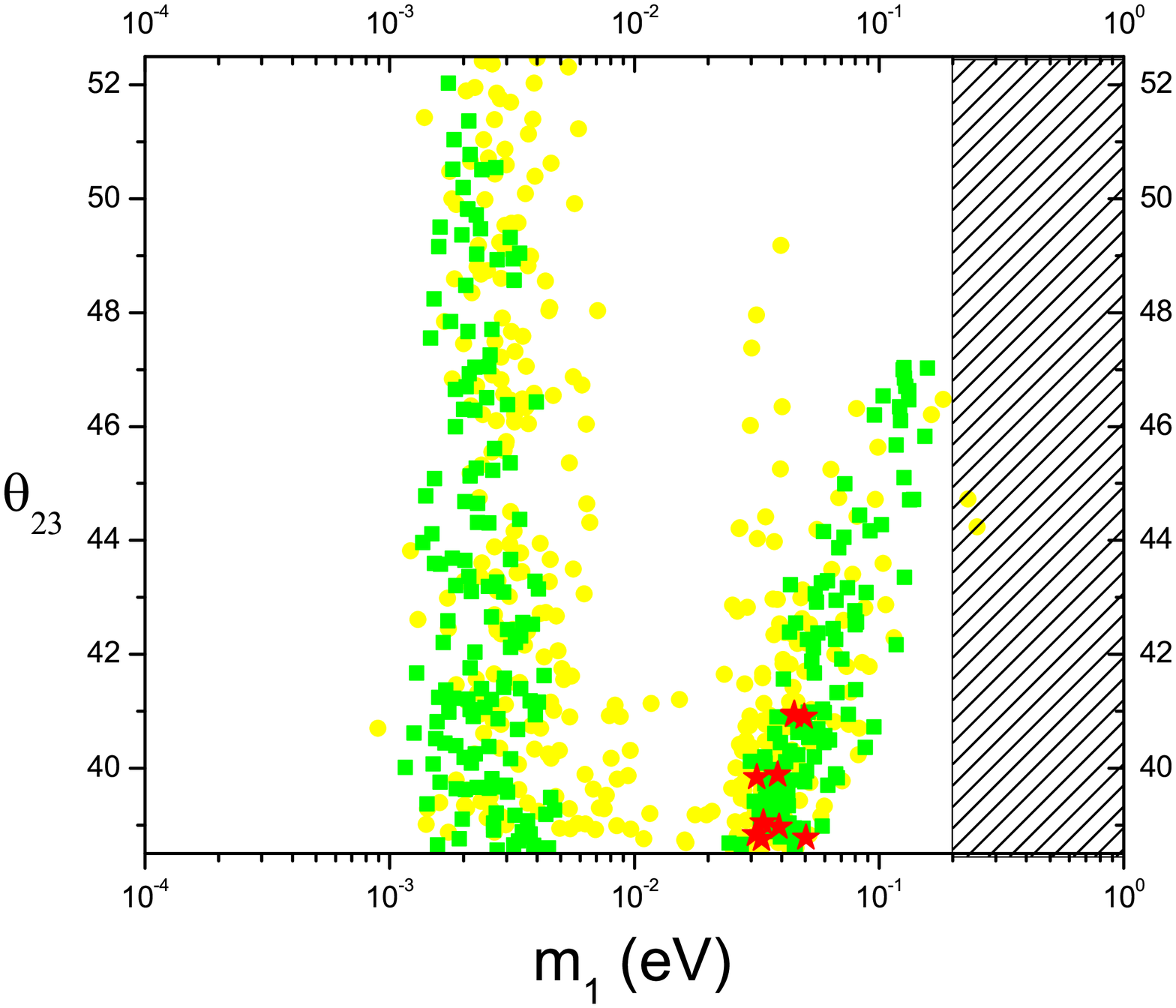,width=55mm,height=55mm}} \\
         \vspace*{15pt}
        \mbox{\epsfig{figure=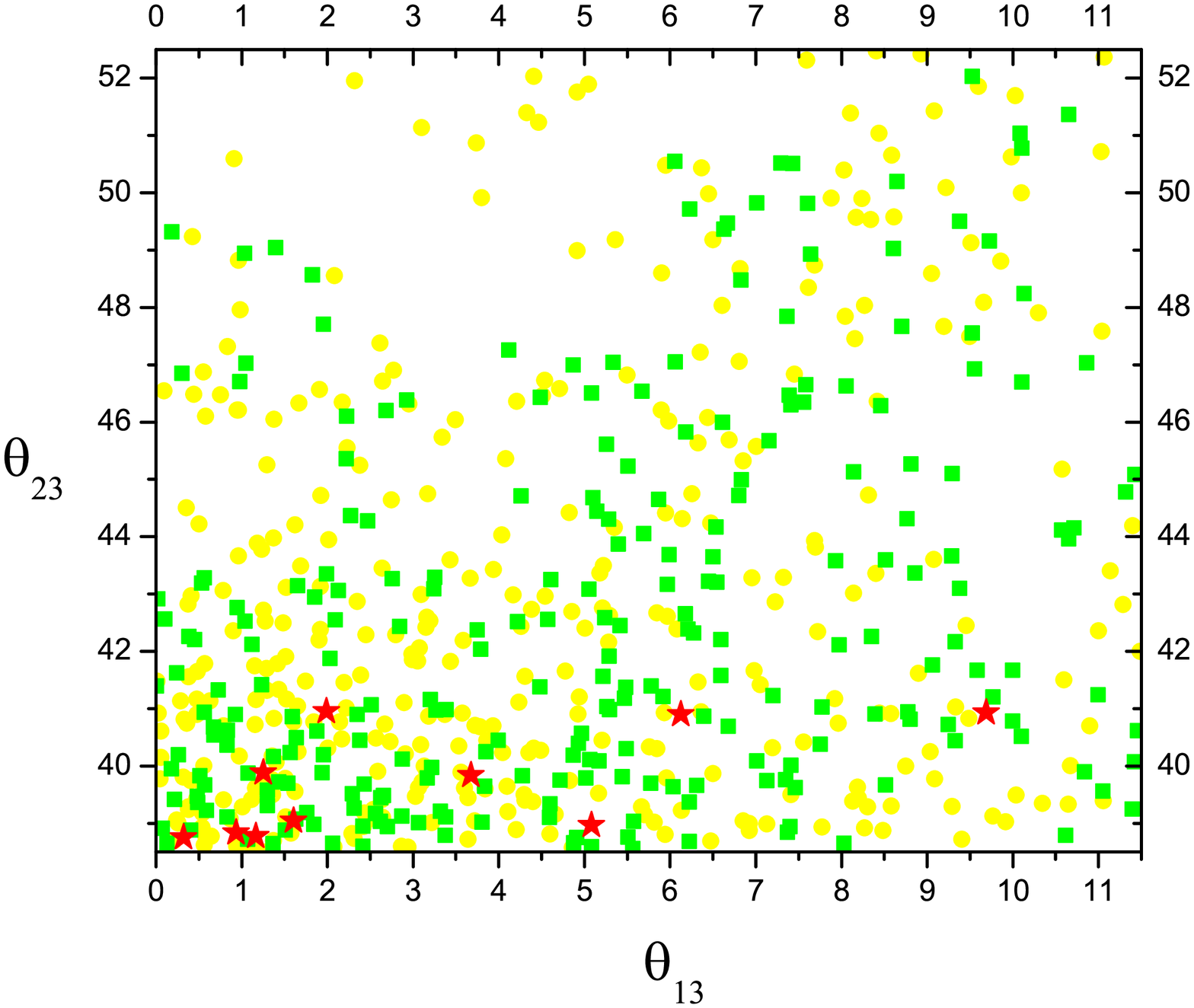,width=55mm,height=55mm}}
        \vspace*{-5pt}
       \mbox{\epsfig{figure=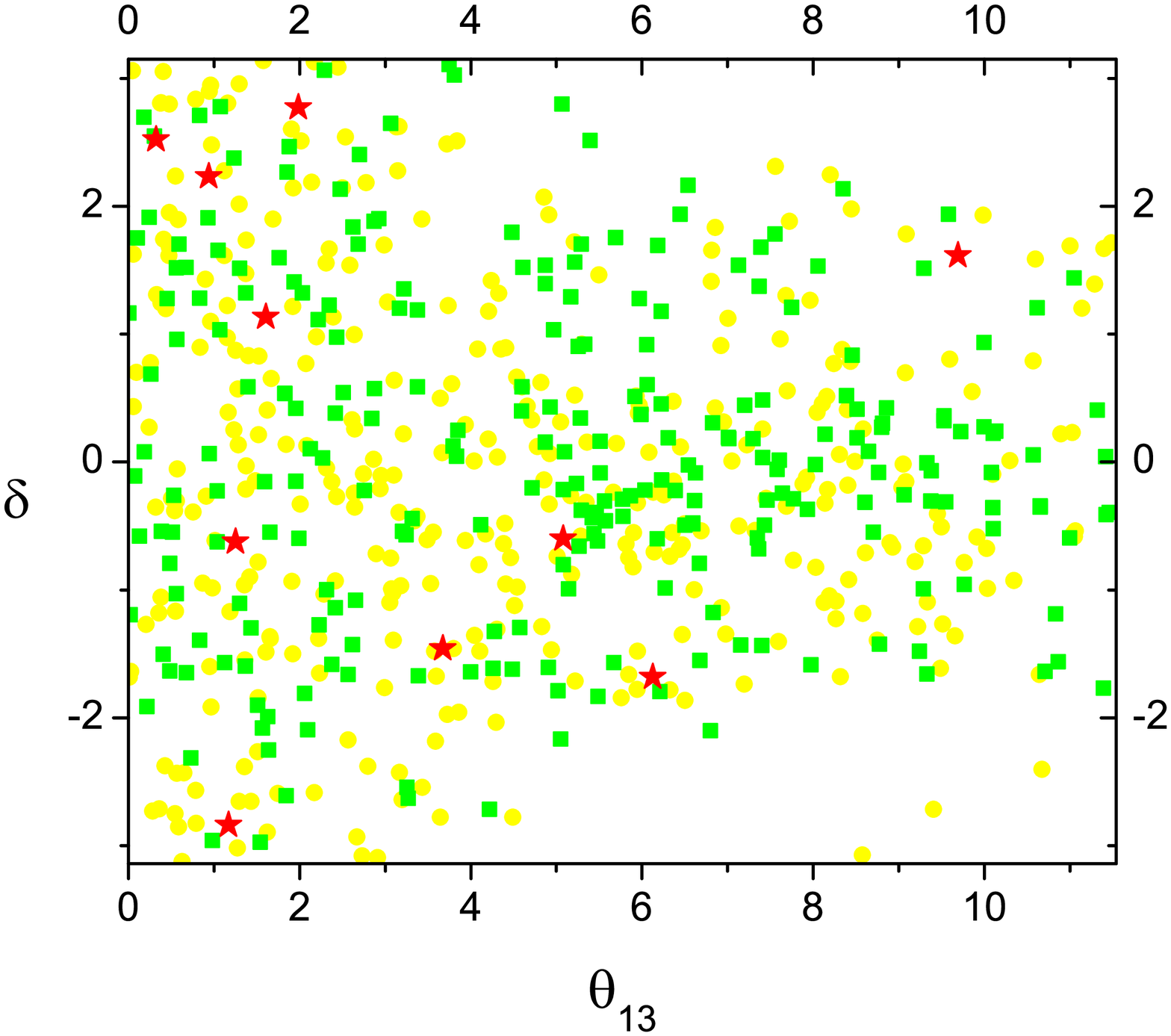,width=55mm,height=55mm}}
\caption{Constraints on some of the low-energy neutrino parameters in the $SO(10)$-inspired
         scenario for normal ordering and $I< V_L < V_{CKM}$~\cite{DiBari:2010ux}. The yellow, green and red points
         correspond respectively to $\a_2=5,4,1$.}
\end{center}
\end{figure}

Together with strongly hierarchical RH neutrino mass patterns,
there exist also  `level crossing' regions where at least two RH neutrino masses 
are arbitrarily close to each other, in a way that  the lightest RH neutrino mass is uplifted and 
$C\!P$ asymmetries are resonantly enhanced to some lever. Also in this case, imposing the successful
leptogenesis condition, one then obtains
conditions, though quite fine tuned ones,  
on the low energy neutrino parameters  \cite{Akhmedov:2003dg}. 

\subsection{Discrete flavour symmetries}

 An account of heavy neutrino flavour effects is also important when leptogenesis is embedded within
theories that try to explain  tribimaximal mixing  for the leptonic
mixing matrix via flavour symmetries. It has been shown in particular that,
if the symmetry is unbroken, then the $C\!P$ asymmetries of the RH neutrinos would exactly
vanish. On the other hand, when the symmetry is broken, for the naturally expected
values of the symmetry  breaking parameters, then the observed
matter-antimatter asymmetry can be successfully reproduced 
\cite{Jenkins:2008rb,Bertuzzo:2009im,Hagedorn:2009jy,aristizabal}.
It is interesting that in a minimal picture based on an $A4$ symmetry, one has a RH neutrino mass spectrum with
$10^{15}\,{\rm GeV} \gtrsim M_3 \gtrsim M_2 \gtrsim  M_1 \gg 10^{12}\,{\rm GeV}$. One has therefore
that all the asymmetry is produced in the unflavoured regime and that the mass spectrum
is only mildly hierarchical (it has actually the same kind of hierarchy as light neutrinos).
At the same time, the small symmetry breaking imposes
a quasi-orthogonality of the three lepton quantum states produced in the RH neutrino
decays. Under these conditions the washout of the asymmetry produced by one RH neutrino species
from the inverse decays of a lighter RH neutrino species is essentially negligible. The final
asymmetry then receives a non-negligible contribution from the decays of all three RH neutrinos species.

\subsection{Supersymmetric models}

Within a supersymmetric vanilla framework the final asymmetry
is only slightly modified compared to the non-supersymmetric calculation \cite{proceedings}.
However, supersymmetry introduces a conceptually important issue: the stringent
lower bound on the reheat temperature, $T_{\rm reh}\gtrsim 10^{9}\,{\rm GeV}$,
is typically marginally compatible with an upper bound
from the avoidance of the gravitino problem $T_{\rm reh}\lesssim 10^{6-10}\,{\rm GeV}$, with the
exact value depending on the parameters of the model \cite{Khlopov:1984pf,Ellis:1984eq,Kawasaki:2008qe}.
It is quite remarkable
that the solution of such an issue inspired an intense research activity on supersymmetric
models able to reconcile minimal leptogenesis and the gravitino problem. Of course, on the
leptogenesis side, some of the discussed extensions beyond the vanilla scenario that relax the RH neutrino
mass bounds also relax the $T_{\rm reh}$ lower bound. However, notice that in the $N_2$ dominated
scenario, while the lower bound on $M_1$ simply disappears, there is still a lower bound
on $T_{\rm reh}$ that is even more stringent, $T_{\rm reh}\gtrsim 6\times 10^{9}\,{\rm GeV}$ \cite{DiBari:2005st}.

As we mentioned already, with flavour effects one has the possibility to relax the lower bound
on $T_{\rm reh}$ if a mild hierarchy in the RH neutrino masses
is allowed together with a mild cancellation in the seesaw formula \cite{bounds}.
However for most models, such as sequential dominated models \cite{King:2003jb},
this solution does not work. A major modification introduced by supersymmetry
is that the critical value of the mass of the decaying RH neutrinos
setting the transition from an unflavoured regime to a two-flavour regime
and from a two-flavour regime to a three flavour regime is enhanced by a factor
$\tan^2\beta$~\cite{Abada:2006fw,Antusch:2006cw}.
This has a practical relevance in the calculation of the asymmetry within supersymmetric models
and it is quite interesting that leptogenesis becomes sensitive to such a relevant
supersymmetric parameter. Recently, a refined analysis,
mainly discussing how the asymmetry is distributed among all particle species,
has shown different subtle effects in the calculation of the final asymmetry
within supersymmetric models finding  corrections below ${\cal O}(1)$ \cite{Fong:2010qh}.

\section{Future prospects for testing leptogenesis}

In 2011 two important  experimental results have been announced that,
if confirmed,  can be interpreted as positive for future tests of leptogenesis.

The first result is the discovery of a non-vanishing
$\theta_{13}\simeq 9^{\circ}$ (cf. Section 2)  confirmed now
by various experiments (both long-baseline and reactor).
An important consequence of the measurement of such a `large'
$\theta_{13}$ is the encouraging prospects for the discovery
of the neutrino mass ordering (either normal or inverted)
in current or near-future neutrino oscillation experiments such as T2K and NO$\nu$A. 
For instance, if the ordering is found
to be inverted in these experiments, we should expect a signal in $0\nu\beta\beta$ experiments
in the next decade. If it is not found, the Majorana nature of neutrinos, and therefore
the seesaw mechanism, would be ruled out.

Moreover, such a large value of $\theta_{13}$ opens the possibility of a measurement of
the neutrino oscillation $C\!P$-violating invariant
$J_{CP}\propto \sin\theta_{13}\,\sin\delta$ during next years.
This would have some direct model-independent consequences.
If a non-vanishing and close-to-maximal value is found ($|\sin\d| \sim 1$),
even the small contribution to the final asymmetry uniquely stemming
from a Dirac phase could be sufficient to reproduce the observed final
asymmetry.
More generally, the presence of $C\!P$ violation at low energies would
certainly support the presence of $C\!P$ violation at high energies as well, since
given a generic theoretical model predicting the neutrino Dirac mass matrix $m_D$,
these are in general both present.

A more practical relevance of such a measurement of $J_{CP}$, even if in the end it
indicates a vanishing value within the experimental error, is that it
will provide an additional constraint on specific models embedding the seesaw,
satisfying successful leptogenesis and able to make predictions on the low energy neutrino parameters.
In this way, the  expected improvements in low-energy neutrino experiments, made easier by
a non-vanishing $\theta_{13}$,  will test the models more and more stringently.

The second important experimental result of 2011 is the hint for
the existence of the Higgs boson reported by the ATLAS and CMS collaborations.
There are at least two reasons why  this hint can be also interpreted as positive for leptogenesis.
First, because the whole  leptogenesis mechanism relies on
the Yukawa coupling between Higgs, lepton and RH neutrino. Second,
because the measurement
of the Higgs boson mass could in future  open opportunities for additional
phenomenological information to be imposed on leptogenesis scenarios,
for example relying on the requirement of Standard Model electroweak 
vacuum stability \cite{giudiceetal}.
In this respect, it is interesting to note that the current value of 125 GeV is compatible 
with reheating temperatures as high as $10^{15}\,{\rm GeV}$
as needed by thermal leptogenesis, and it is also compatible with the requirement that the
Yukawa couplings of the RH neutrinos do not destabilise the Higgs potential  
when RH neutrino masses $M_i$, assumed to be quasi-degenerate, are smaller than $10^{14}$~GeV.

What are other possible future experimental developments
that could further support the idea of leptogenesis?
An improved information from absolute neutrino mass scale experiments,
both on the sum of the neutrino masses from cosmology and on $m_{ee}$
from $0\n\b\b$ experiments could be crucial. For instance,
if cosmology provides a measurement of the neutrino mass
$0.01\,{\rm eV} \lesssim m_1 \lesssim 0.2\,{\rm eV}$
in the next few years, as it is reasonable to expect, then a positive signal
in $0\n\b\b$ experiments must be found, otherwise Majorana neutrinos and the seesaw
mechanism will be disfavored.
On the other hand, in case of a positive signal in $0\n\b\b$ experiments, we
will be able to say that minimal leptogenesis with hierarchical heavy neutrino
masses works in a optimal neutrino mass window
$10^{-3}\,{\rm eV}\lesssim m_1 \lesssim 0.1\,{\rm eV}$ \cite{window,bounds,problem}, where
independence of the initial conditions is more easily obtained and where we know that successful
leptogenesis can be safely obtained from existing calculations using Boltzmann equations.
Moreover, a determination of the allowed region in the plane $m_{ee}$--$\sum_i\,m_i$
could provide an additional test of specific leptogenesis scenarios, such as, for example, 
$SO(10)$-inspired scenarios.

In conclusion, we are living in an exciting time where new experimental information
is coming, which is providing and will continue to provide crucial tests for new
physics models, in particular leptogenesis, in the near future.

\section*{Acknowledgements}

We wish to thank A.~Abada, W.~Buchm\"{u}ller, S.~Davidson, M.~Drewes, T.~Hambye, 
D.A.~Jones, S.F.~King, L.~Marzola, A.~Pilaftsis for useful comments and discussions.
We also wish to thank A.~Hohenegger for enlightening discussions about the closed-time-path formalism.
PDB also wishes the   Laboratoire de Physique Th\'{e}orique, Universit\'{e} de Paris-Sud 11 (Orsay) 
for the warm hospitality during the completion of this work.
SB acknowledges support from the Swiss National Science Foundation, under the Ambizione grant
PZ00P2\_136947.  PDB acknowledges financial support  from the NExT/SEPnet Institute, 
from the STFC Rolling Grant ST/G000557/1 and from the  EU FP7  ITN INVISIBLES 
(Marie Curie Actions, PITN- GA-2011- 289442).

\section*{References}

\end{document}